\documentclass[twocolumn]{aastex62}
\shorttitle{Solar Wind Abundances}
\shortauthors{Laming et al.}

\begin{document}

\title{Element Abundances: A New Diagnostic for the Solar Wind}


\correspondingauthor{J. Martin Laming}\email{laming@nrl.navy.mil}

\author[0000-0002-3362-7040]{J. Martin Laming}
\affiliation{Space Science Division, Code 7684, Naval Research Laboratory,
Washington DC 20375, USA}

\author[0000-0002-8164-5948]{Angelos Vourlidas}
\affiliation{Johns Hopkins University Applied Physics Laboratory, Laurel MD
20723, USA}

\author{Clarence Korendyke}
\affiliation{Space Science Division, Code 7686, Naval Research Laboratory,
Washington DC 20375, USA}

\author{Damien Chua}
\affiliation{Space Science Division, Code 7686, Naval Research Laboratory,
Washington DC 20375, USA}

\author[0000-0002-3699-3134]{Steven R. Cranmer}
\affil{Department of Astrophysical and Planetary Sciences, Laboratory for
Atmospheric and Space Physics, University of Colorado, Boulder, CO 80309, USA}

\author[0000-0002-8747-4772]{Yuan-Kuen Ko}
\affiliation{Space Science Division, Code 7684, Naval Research Laboratory,
Washington DC 20375, USA}

\author{Natsuha Kuroda}
\affiliation{University Corporation for Atmospheric Research (UCAR), Boulder
CO 80307, USA, and Space Science Division, Code 7684, Naval Research
Laboratory, Washington DC 20375, USA}

\author{Elena Provornikova}
\affiliation{Johns Hopkins University Applied Physics Laboratory, Laurel MD
20723, USA}

\author[0000-0002-7868-1622]{John C. Raymond}
\affiliation{Smithsonian Astrophysical Observatory, Cambridge MA 02138, USA}

\author{Nour-Eddine Raouafi}
\affiliation{Johns Hopkins University Applied Physics Laboratory, Laurel MD
20723, USA}

\author{Leonard Strachan}
\affiliation{Space Science Division, Code 7684, Naval Research Laboratory,
Washington DC 20375, USA}

\author{Samuel Tun-Beltran}
\affiliation{Space Science Division, Code 7684, Naval Research Laboratory,
Washington DC 20375, USA}

\author{Micah Weberg}
\affiliation{NRL/NRC Research Associate, Space Science Division, Code 7684,
Naval Research Laboratory, Washington DC 20375, USA}

\author[0000-0002-4998-0893]{Brian E. Wood}
\affiliation{Space Science Division, Code 7685, Naval Research Laboratory,
Washington DC 20375, USA}


\begin{abstract}
We examine the different element abundances exhibited by the closed loop
solar corona and the slow speed solar wind. Both are subject to the First
Ionization Potential (FIP) Effect, the enhancement in coronal abundance of
elements with FIP below 10 eV (e.g. Mg, Si, Fe) with respect to high FIP
elements (e.g. O, Ne, Ar), but with subtle differences. Intermediate
elements, S, P, and C, with FIP just above 10 eV, behave as high FIP
elements in closed loops, but are fractionated more like low FIP elements
in the solar wind. On the basis of FIP fractionation by the ponderomotive
force in the chromosphere, we discuss fractionation scenarios where this
difference might originate. Fractionation low in the chromosphere where
hydrogen is neutral enhances the S, P and C abundances. This arises with
nonresonant waves, which are ubiquitous in open field regions, and is also
stronger with torsional Alfv\'en waves, as opposed to shear (i.e. planar)
waves. We discuss the bearing these findings have on models of interchange
reconnection as the source of the slow speed solar wind. The outflowing
solar wind must ultimately be a mixture of the plasma in the originally
open and closed fields, and the proportions and degree of mixing should
depend on details of the reconnection process. We also describe novel
diagnostics in ultraviolet and extreme ultraviolet spectroscopy now
available with these new insights, with the prospect of investigating slow
speed solar wind origins and the contribution of interchange reconnection
by remote sensing.

\end{abstract}


\section{Introduction}
The prediction of the existence of the solar wind \citep{parker58} must rank
as one of the key theoretical insights in the history of heliophysics. Since
its discovery \citep{gringauz60,neugebauer62}, Parker's original concept of a
wind driven by thermal pressure in a corona heated by magnetohydrodynamic
(MHD) waves \citep{parker63} has been slightly modified to a scenario where
the MHD waves drive the wind directly
\citep[e.g.][]{belcher71,isenberg82,ofman10}. The fast solar wind is
established to emerge from coronal holes, open field regions where plasma
emerges directly from the solar chromosphere into the wind, and exhibits
largely unbalanced Alfv\'enic turbulence \citep{bruno13,ko18}. By contrast
the slow solar wind, which shows strong chemical fractionation effects in its
composition and more balanced (or high cross-helicity) turbulence, is
frequently believed to originate in closed coronal loops where the
fractionation occurs \citep[e.g.][]{antiochos11}, before being released into
the solar wind by interchange reconnection with surrounding open field, as
well as possibly coming directly from open field like the fast solar wind
\citep{cranmer07}.

%In this paper we advocate a new view of the solar upper atmosphere and wind,
%and discuss the instrumentation desirable for further investigations.
Solar wind acceleration and composition depend on processes at three
transition layers in the solar upper atmosphere. The first, usually located
in the low chromosphere, is where the pressure changes from being thermally
dominated to being magnetically dominated. In this region the sound speed and
Alfv\'en speed are equal, and a number of processes involving wave mode
conversion and other wave-wave interactions can occur. This is where a
significant fraction of the MHD waves that eventually accelerate the solar
wind are generated from motions ultimately deriving from solar convection.
The second transition layer appears higher up in the chromosphere, where
largely neutral gas gives way to the ionized plasma that ends up as the solar
corona and wind. This transition gives rise to strong density gradients and
associated wave reflection and refraction. Alfv\'en waves interacting with
this density gradient generate the ponderomotive force. This combines the
effect of the wave pressure gradient and the force on the plasma wave due to
wave reflection and refraction. Since the waves are fundamentally magnetic in
character, only ions see this force, and ion-neutral separation is the
result, giving rise to element fractionation in the upper atmosphere known as
the First Ionization Potential (FIP) Effect. This abundance anomaly has been
seen in the solar corona and wind for over fifty years
\citep[e.g.][]{pottasch63}, and can be seen to offer a key observable for
wave processes that until now has remained largely unexploited.

\begin{figure}[t]
\centerline{\includegraphics[scale=0.45]{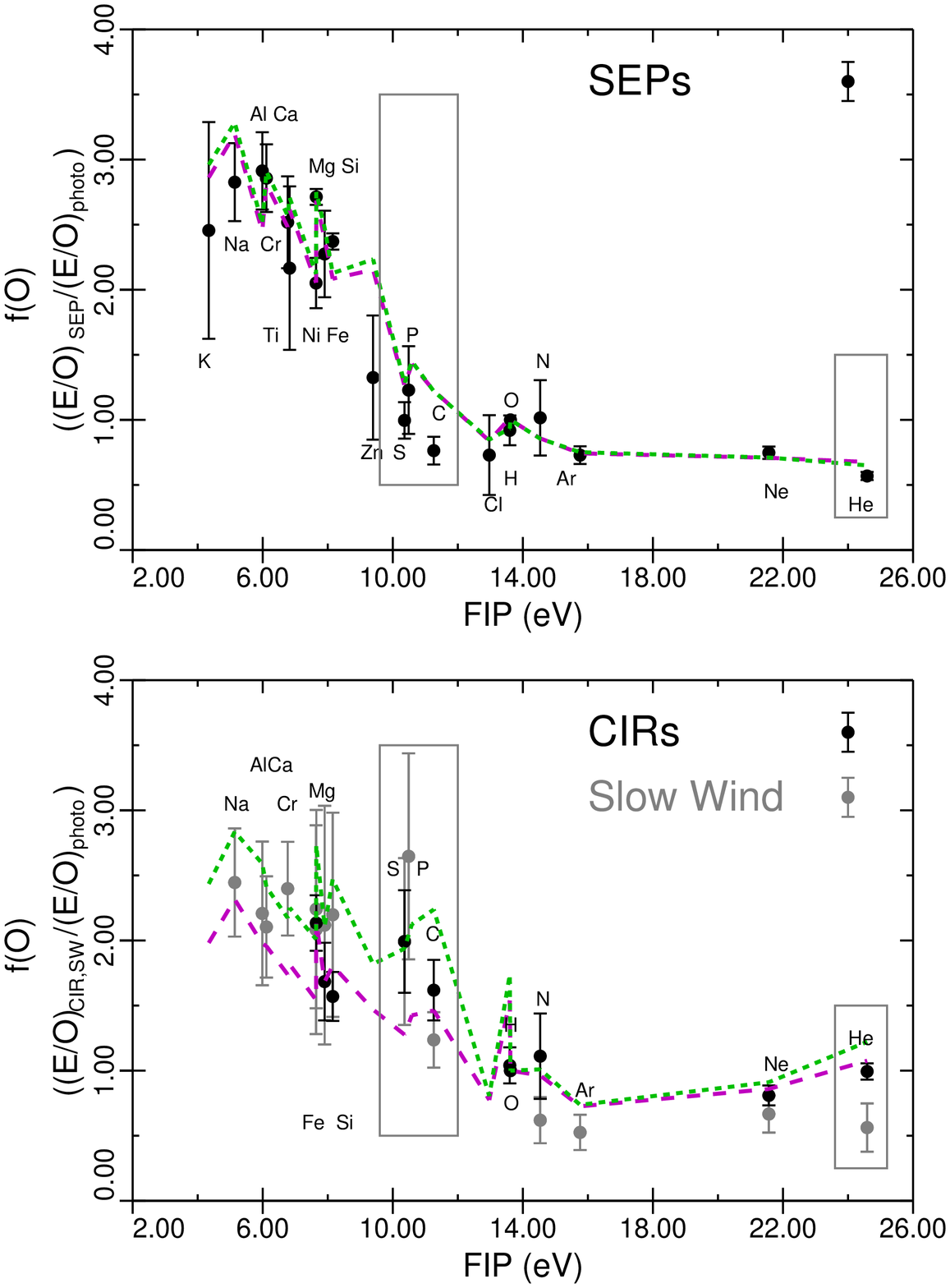}}
\centerline{\includegraphics[scale=0.45]{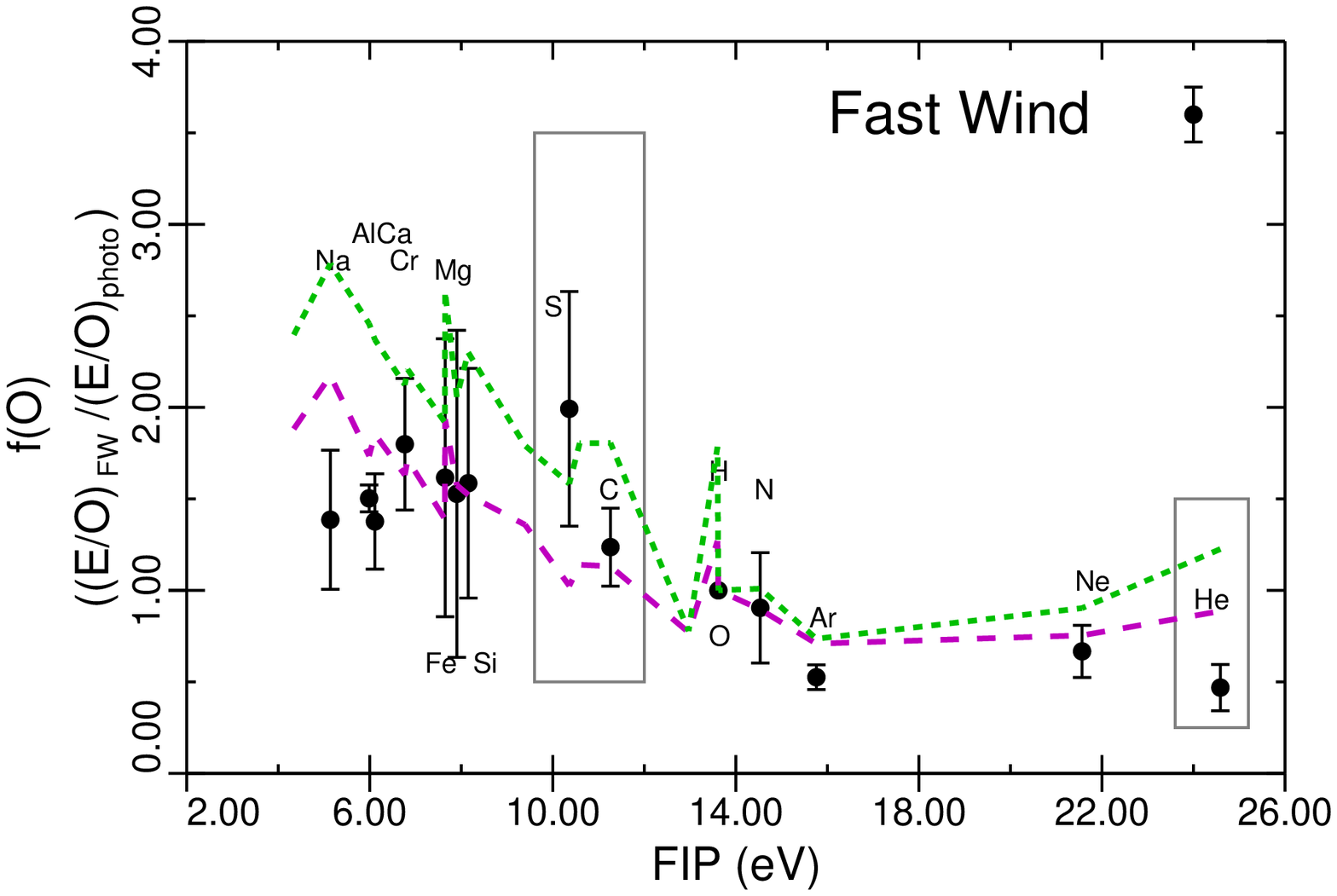}}
%\plotone{f1.eps}
\caption{\label{fig1} Variation of the FIP Fractionation. Top: SEP fractionations relative to O shown as black circles
with error bars from Table 1 in Reames (2018). Model calculations for a closed
coronal loop are shown as a result of ponderomotive
FIP fractionation by shear (magenta) and torsional (green) Alfv\'en waves and mass dependent adiabatic invariant
conservation (see sections 2 and 3). Middle: CIR energetic particle fractionations (black circles, Reames 2018) and
slow speed solar wind (dark gray, Bochsler 2009) relative to O.
Models are for an open field with $B=30$G in the corona. Bottom: Fast solar wind
fractionations (Bochsler 2009) compared
with open field models with coronal field of 10 G. Boxes highlight the S, P, C, and He
fractionations that are especially variable.
}
\end{figure}

The third transition layer, and arguably the hardest to understand in a
quantitative theoretical manner, is the evolution of the solar plasma from a
fluid to a collisionless plasma dominated by kinetic effects. This happens
where the ion-proton collision rate becomes slower than the solar wind
expansion rate, $v_w\left(r\right)/r$, where $v_w\left(r\right)$ is the solar
wind speed and $r$ is the heliocentric radius. With the ion-proton collision
rate at freeze-in given by
\begin{equation}
\nu _{ip}={4\pi e^4\ln\Lambda\over m_p}{n\over k_{\rm B}T}{Z^2\over A+1}
{4\over 3\sqrt{\pi}}\sqrt{{A\over A+1}{m_p\over 2k_{\rm B}T}}\sim
{v_w\left(r\right)\over r}
\end{equation}
for an ion of charge $Z$, mass $m_pA$ where $m_p$ is the proton mass, plasma
parameter $\ln\Lambda\simeq 20$, and all other symbols have their usual
meanings, this transition is expected to occur at a plasma density $n\sim
10^6$ cm$^{-3}$, which corresponds to a radius $r\sim 1.5 R_{\sun}$ where
$v_w\sim 100$ km s$^{-1}$ in the slow wind. In this region the density varies
most strongly, and largely controls this transition. Obviously, ions of
different elements will make this transition at various radii, leading to a
much less ``clean'' transition than either of the first two. But this
transition is crucially important to the wave-driven acceleration of the
solar wind. Ions can only be accelerated by the absorption of ion cyclotron
waves once they are decoupled from fluid motions \citep{cranmer99,
miralles01}. Different ions will accelerate at different rates depending on
where exactly they decouple and on how much MHD wave energy is available to
them at their ion cyclotron resonant frequency.

The behavior of waves and how they interact with these three transition
layers is crucial to the acceleration and elemental composition of the solar
wind. The varieties of fractionation that are routinely exhibited by the
solar corona and wind are shown in Figure 1, replotted from data given in
Table 1 of \citet{reames18}. The top panel shows element fractionations for
Solar Energetic Particles (SEPs) relative to the O abundance of
\citet{caffau11}, given as black circles with error bars, together with model
calculation designed to match abundances determined remotely by spectroscopy
of a closed coronal loop, with an assumed coronal magnetic field of 30G for
shear (magenta dashed curve) and torsional (green dashed curve) Alfv\'en
waves. The calculation is described in detail below, but for now we point out
that both in observations and the model, S, P and C behave mainly as high FIP
elements, being fractionated by an insignificant amount. \citet{reames18}
points out that this correspondence in element abundances between SEPs and
closed coronal loops means that the particles that end up being accelerated
in shock waves must have an origin in the closed loop solar corona, and
cannot be swept up out of the ambient solar wind, as previously argued
elsewhere. \citet[and references therein]{laming13} reach the same conclusion
on somewhat different grounds.

The middle panel shows similar measurements for accelerated particles
measured in Corotating Interaction Regions (CIRs, black symbols) and slow
solar wind (dark gray symbols). In contrast with the case above, in CIRs, the
accelerated particles are swept up directly from the solar wind, hence the
correspondence between the sets of observations. The models show
fractionations in open field with a magnetic field of 30G at the top of the
chromosphere for shear and torsional Alfv\'en waves as before. A key
difference, picked up by both model curves but especially by the torsional
Alfv\'en waves, is that S, P and C now behave more like low FIP elements.
This difference in behavior between SEPs and solar wind had been visible
previously in SEPs \citep{ko13} and solar flares
\citep{sylwester08,sylwester12}, compared with solar wind observations
\citep{giammanco07a,giammanco07b,giammanco08,reisenfeld07}. It is also
displayed in Figure 1 of \citet{schmelz12}, and first commented, to our
knowledge, by \citet{rakowski12}. As we argue further below, this difference
in fractionation pattern is crucial to understanding slow solar wind origin,
and the processes such as interchange reconnection that form it.

Finally, for completeness, in the bottom panel we show results for the fast
solar wind, together with models (again for shear and torsional waves) for
open field regions with coronal fields of and 10 G. With the exception of S
which has large error bars,\citep[but is measured lower by][]{gloeckler07},
shear Alfv\'en waves (the magenta curve) are clearly favored by the model
(more details given in section 2), while torsional waves better reproduce the
slow solar wind abundances, especially S, P, and C.

Although interchange reconnection was originally introduced as a means of
releasing FIP fractionated material in closed loops into the solar wind, we
are finding that it is also important as a source of torsional Alfv\'en
waves, which we discuss further below. Thus plasma fractionation on open
field lines may be qualitatively different in coronal holes away from closed
fields or in active regions close to closed field regions.

The differences between these panels suggest that possibilities exist for
diagnosing the origin of the solar wind in terms of the magnetic geometry of
the structure(s) from which it emanates in terms of the microphysics as
embodied by the element abundances. In section 2 of this paper, we give a
more detailed discussion of the origins of the FIP fractionation and how the
variations in fractionation may be related to wave properties in different
magnetic structures. Section 3 gives model results, while section 4
summarizes other possible mechanisms of fractionation. Section 5 outlines an
observational approach to validate some of these hypotheses, and section 6
concludes.

\begin{figure}[t]
\centerline{\includegraphics[scale=0.45]{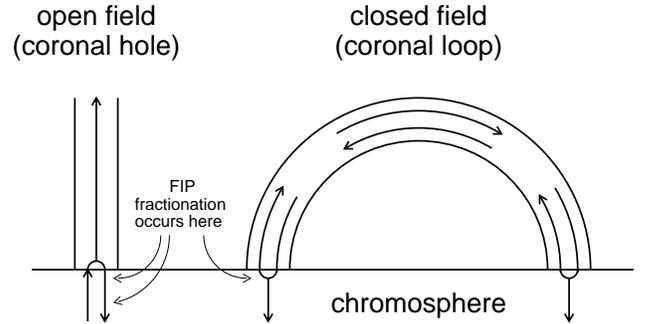}}
%\plotone{f2.eps}
\caption{\label{fig2} Schematic showing FIP fractionation in open and closed field
regions. In open field, waves impinge on footpoints from below, but in closed field
wave generation within the coronal loop dominates.}
\end{figure}

\section{FIP Fractionation}
\subsection{Open Versus Closed Field}
We describe in more detail the calculations producing the model results in
Fig. 1. We begin with the two simple scenarios shown schematically in Fig. 2,
an open field region and a closed coronal loop, which serve as the basic
models for fast solar wind and coronal or SEP abundances respectively. In
open field regions, waves deriving from convection within the solar envelope
propagate upwards to the footpoint, and either enter the coronal hole or are
reflected back down again. These waves entering the coronal hole ultimately
drive the solar wind outflow. Typically the periods of these waves (three or
five minutes) are too long for resonance with a closed coronal loop, and so
in this case they are generally reflected back downwards when encountering
the footpoints of such loops. Resonant waves are most plausibly excited
within the coronal loop itself, most likely as a byproduct of the
mechanism(s) that heat the corona \citep{dahlburg16,tarr17}. In open field
regions, such a resonance does not exist, and only waves propagating up from
footpoints are possible. \citet{cranmer07} was able to show that the MHD
turbulence in an open slow-solar-wind flux tube could have some low-FIP
abundance enhancement (i.e., the Fe/O ratio) without the need for closed
loops to undergo interchange reconnection. Hence the slow solar wind
composition is most likely a combination of the compositions arising from the
two scenarios, as a closed loop interchange reconnects
\citep[e.g.][]{lynch14,higginson17} with neighboring open field to release
its plasma into the solar wind. This has been recently discussed in terms of
the evolution of the Separatrix-Web \citep[S-Web;][]{antiochos11}, the
network of quasi-separatrix layers formed by open field corridors within
otherwise closed field regions.

Interchange reconnection is also important in exciting torsional Alfv\'en
waves. \citet{lynch14,higginson17} report simulations showing a large scale
torsional Alfv\'en wave arising as open field reconnects with a twisted
closed loop, with the twist being transferred to the resulting open field.
The twist excites torsional waves as it relaxes. In the models presented
above ponderomotive force from torsional waves seems to be important in
fractionating S/O to the levels seen in the slow speed solar wind, and the
torsional wave amplitudes seen in the simulation of \citet{lynch14} and in
the observations of \citet{tiwari18} are consistent with our models. It is
possible that this wave is not involved in the solar wind acceleration.
\citet{vasheghani12} argue that these waves do not couple to other modes or
to each other as well as shear Alfv\'en waves, meaning that any turbulent
cascade will be less efficient in producing waves in the ion-cyclotron range
that can resonant with solar wind ions. But conversely they might better
survive propagation through the chromosphere to fractionate the plasma. In
this way the fast wind is also different; the (shear) waves that fractionate
the plasma are also taken to be the waves that reflect, cascade and
ultimately accelerate the solar wind.

\subsection{Ponderomotive Ion-Neutral Separation} \citet{laming04} introduced the idea
that ion-neutral separation in the chromosphere arises as a result of the
ponderomotive force. This force arises as a result of Alfv\'en or fast mode
(collectively known as ``Alfv\'enic'' when close to parallel propagation)
waves propagating through or reflecting from the solar chromosphere. In the
absence of wave reflection or refraction (the WKB approximation), the
ponderomotive force is just the negative wave pressure gradient. However, in
the presence of wave reflection or refraction, the wave particle interaction
is mediated through the refractive index of the plasma, with the result that
MHD waves and ions are attracted to each other (the opposite of the negative
wave pressure gradient). A general form for the instantaneous ponderomotive
acceleration, $a$, experienced by an ion is \citep[see e.g. the appendix
of][]{laming16}
\begin{equation}
a={c^2\over 2}{\partial\over\partial z}\left(\delta E^2\over B^2\right)
\end{equation}
where $\delta E$ is the wave electric field, $B$ the ambient magnetic field,
$c$ the speed of light, and $z$ is a coordinate along the magnetic field.

The element fractionation by the ponderomotive force is calculated from
momentum equations for ions and neutrals in a background of protons and
neutral hydrogen. The ratios, $f_k$, of densities $\rho _k$ for element $k$
at upper and lower boundaries of the fractionation region $z_u$ and $z_l$
respectively, is given by the equation \citep{laming16}
\begin{eqnarray}
\nonumber f_k&=&{\rho _k\left(z_u\right)\over\rho _k\left(z_l\right)}\\ &=&\exp\left\{
\int _{z_l}^{z_u}{2\xi _ka\nu _{kn}/\left[\xi _k\nu
_{kn} +\left(1-\xi _k\right)\nu _{ki}\right]\over 2k_{\rm B}T/m_k+v_{||,osc}^2+2u_k^2}dz\right\},
\end{eqnarray}
where $\xi _k$ is the element ionization fraction, $\nu _{ki}$ and $\nu
_{kn}$ are collision frequencies of ions and neutrals with the background gas
\citep[mainly hydrogen and protons, given by formulae in][]{laming04},
$k_{\rm B}T/m_k \left( =v_z^2\right)$ represents the square of the element
thermal velocity along the $z$-direction, $u_k$ is the upward flow speed and
$v_{||,osc}$ a longitudinal oscillatory speed, corresponding to upward and
downward propagating sound waves. Because $\nu _{ki}>>\nu _{kn}$ in the
fractionation region at the top of the chromosphere, small departures of $\xi
_k$ from unity can result in large decreases in the fractionation. This
feature is important in suppressing the fractionation of S, P, and C at the
top of the chromosphere, while allowing it lower down where the H is neutral,
giving rise to the different fractionation of these elements in the various
panels of Fig. 1.

The specification of $v_{||,osc}$ is outlined in the next subsection. Here we
describe the implementation of some important approximations near the plasma
$\beta =1$ layer. When $v_{||,osc}$ is greater than the local Alfv\'en speed,
all fractionation is assumed to cease. We argue that the sound waves will
excite counter-propagating Alfv\'en waves which can then cascade to
microscopic scales mixing the plasma at a rate much faster than it can be
fractionated. In general, $v_{||,osc}$ has contributions from upward
propagating sound waves excited by solar convection, and sound wave excited
in the chromosphere by the Alfv\'en wave driver \citep[e.g.][]{arber16} by
the modulational instability \citep[sometimes known as parametric excitation
e.g.][]{landau76}. Similar arguments restrict the fractionation to the plasma
$\beta < 1$ region of the solar atmosphere. Here sound waves can also decay
directly to counter-propagating Alfv\'en waves which can again cause mixing
after cascading to microscopic scales.

\subsection{Chromospheric Model and Wave Fields}
We take the chromospheric model of \citet{avrett08} for temperature, density
and electron density profiles, combined with a force free magnetic field
calculated from formulae given by \citet{athay81}. This captures the behavior
low in the chromosphere as the magnetic field decreases with height near the
plasma $\beta =1$ layer, and is constant with height above this region. The
temperature and density profiles are shown in the top left panel of Fig. 3.
The hydrogen ionization balance dominating the electron density is shown as
the thick line on the top right panel of Fig. 3. The degree of ionization
inferred observationally is higher than equilibrium at the local density and
temperature would suggest, presumably due to the passage of shock waves that
elevate the ionization fraction on timescales faster than that associated
with electron-proton recombination \citep{carlsson02}. Ionization balances
for other elements are calculated here using the local temperature, density,
and radiation field. This comprises coronal radiation from above, taken from
\citet{vernazza78}, absorbed progressively in the chromosphere, and trapped
chromospheric Lyman $\alpha$ photons. The coronal spectrum varies from
coronal holes to active regions, introducing small variations in the
ionization fraction of minor ions, most visible in the high FIP elements.
Atomic data are taken from \citet{verner96} for photoionization cross
sections, and from \citet{mazzotta98} for collisional rates. We are only
concerned with neutral atoms and singly charged ions so subsequent
refinements to dielectronic recombination rates as considered in
\citet{bryans09} are largely unimportant. We do however include the effects
of electron density on the dielectronic recombination, following
\citet{nikolic13}. Uncertainties in the ionization balance are probably
dominated by the underlying chromospheric model and the assumed coronal
ionizing spectrum rather than by atomic data deficiencies.

\begin{figure*}[t]
\centerline{\includegraphics[scale=1.0]{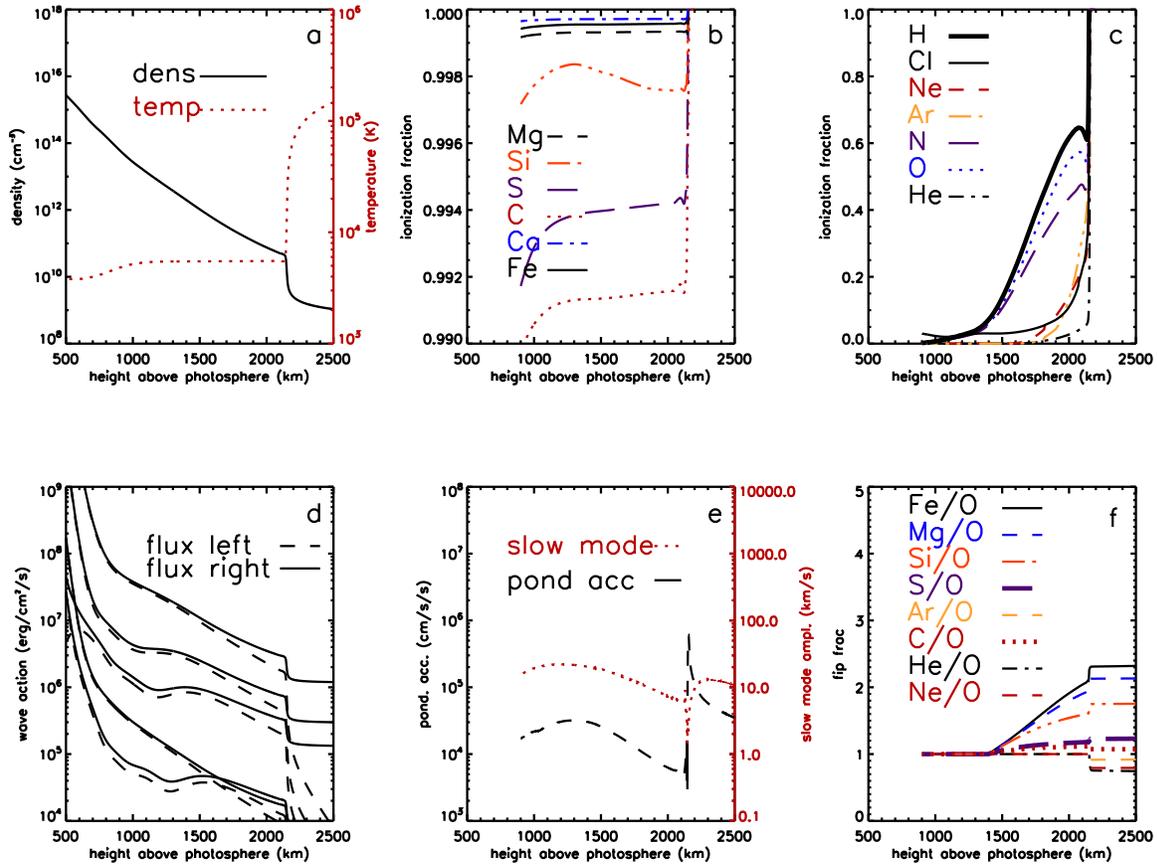}}
%\plotone{f3.ps}
\caption{\label{fig3}The chromospheric model for fast wind from an open field region. (a) shows the
density and temperature structure of the chromosphere. (b) shows chromospheric
ionization fractions for low FIP elements and (c) for high FIP elements.
(d) shows the wave energy fluxes in each direction for the five waves
in the open field model. (e) shows the ponderomotive acceleration (solid line)
and the
amplitude of slow mode waves induced by the Alfv\'en wave driver. (f) shows the
fractionations resulting for selected elements relative to O, S and C are shown
with thicker lines. Gas pressure and magnetic field
pressure are equal at about 1,000 km, magnetic field pressure dominating at higher
altitudes.}
\end{figure*}

Chromospheric acoustic waves are introduced to match simulations and data
analysis in \citet{heggland11} and \citet{carlsson15}. Acoustic waves with a
flux of $10^8$ ergs cm$^{-2}$s$^{-1}$ propagate upwards through the
chromosphere with their amplitude increasing as the density decreases in
accordance with the WKB approximation, until the amplitude reaches the local
sound speed. At this point we stop the amplitude growth, arguing that the
excess energy is lost to the wave by radiation and conduction, principally
cooling by Lyman $\alpha$ with a timescale of the order of seconds. Lower
down the cooling is dominated by H$^-$ with a timescale of order minutes
\citep{ayres96}.

The Alfv\'en waves are modeled using transport equations given by
\citet{cranmer05} and \citet{laming15},
\begin{equation}
{\partial I_{\pm}\over\partial t}+\left(u\mp V_A\right){\partial I_{\pm}\over\partial z}
=\left(u\pm V_A\right)\left({I_{\pm}\over 4L} +{I_{\mp}\over 2L_A}\right),
\end{equation}
where ${\bf I}_{\pm}=\delta {\bf v}\pm \delta {\bf B}/\sqrt{4\pi\rho}$ are
the Els\"asser variables representing waves propagating in the $\mp$
z-directions. The Alfv\'en wave spectrum in the coronal hole model is taken
from from \citet{cranmer07}. We specify five waves to match the peaks in the
theoretical spectrum, and start the integration at an altitude of 500,000 km,
where the outgoing waves dominate \citep{cranmer05}. At an altitude of 1,000
km in the coronal hole, where the sound speed and Alfv\'en speeds are equal,
the Alfv\'en wave solution corresponds to an energy flux of $\sim 4\times
10^7$ ergs cm$^{-2}$s$^{-1}$, comparable to but slightly less than the upward
acoustic wave energy flux that generates these waves.

In the closed field model, we assume coronal waves only, which are taken to
be the loop resonant mode as in \citet{laming12,laming16} and
\citet{rakowski12}. The amplitude is adjusted to give a best match with
observed FIP fractionations, and is typically $\sim 50$ km s$^{-1}$.
Simulations of coronal heating show that waves of this amplitude are indeed
produced as a ``by-product'' of the heating mechanism
\citep{dahlburg16,tarr17}.

\begin{figure*}[t]
\centerline{\includegraphics[scale=1.0]{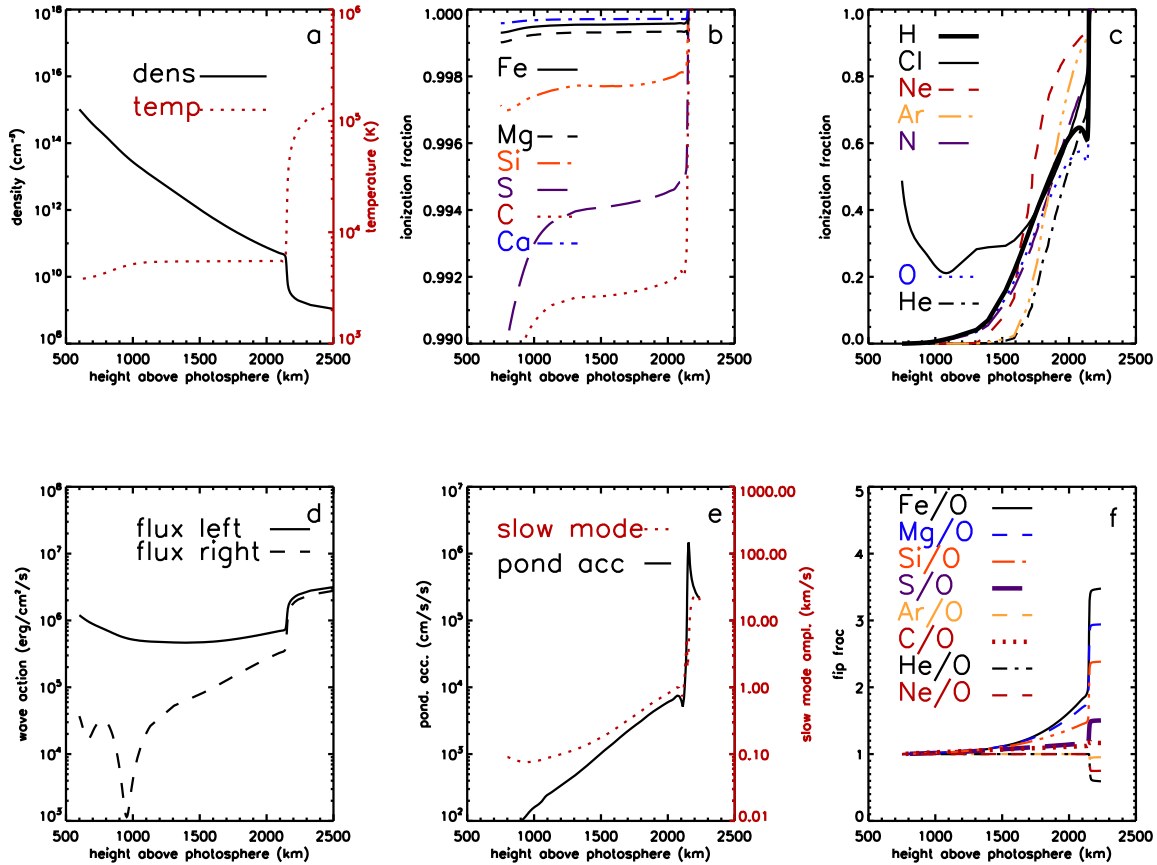}}
%\plotone{f4.ps}
\caption{\label{fig4}The chromospheric model for a closed field region. (a) shows the
density and temperature structure of the chromosphere. (b) shows chromospheric
ionization fractions for low FIP elements and (c) for high FIP elements.
(d) shows the wave energy fluxes in each direction for the resonant wave
in the closed field model. (e) shows the ponderomotive acceleration (solid line)
and the
amplitude of slow mode waves induced by the Alfv\'en wave driver. (f) shows the
fractionations resulting for selected elements relative to O, S and c are shown
with thicker lines. Gas pressure and magnetic field
pressure are equal at about 750 km, magnetic field pressure dominating at higher altitudes.}
\end{figure*}

At high Alfv\'en wave energy fluxes, the ponderomotive force will modify the
structure of the chromosphere itself. We estimate when this will occur as
follows. The expression for the ponderomotive acceleration can be modified to
\citep{laming15}
\begin{equation}
a={a_0\over 1+\left(\xi _h/4\right)\left(\nu _{eff}/\nu _{hi}\right)
\left(\sum _{waves}\delta v^2/v_h^2\right)}
\end{equation}
where $\nu _{eff}=\nu _{hi}\nu _{hn}/\left(\xi _h\nu _{hn}+\left(1-\xi
_h\right)\nu _{hi}\right)$ is the effective collision frequency of element
$h$ (in this case hydrogen) in terms of its collision frequencies when
ionized $\nu _{hi}$, and when neutral $\nu _{hn}$, and $v_h^2=k_{\rm B}T/m_h
+v_{||,osc}^2/2+u_k^2$ is the square of the hydrogen speed, in terms of its
thermal speed, the amplitude of slow mode waves propagating through the
chromosphere, and the flow speed in the chromospheric model. Since the
ponderomotive force separates ions from neutrals, its effect on the
background plasma to smooth out density gradients depends on the coupling
between ionized and neutral hydrogen, and is strongest in regions where
hydrogen is fully ionized ($\xi =1$), and becomes significant at
ponderomotive accelerations above about $10^6$ cm s$^{-2}$, possibly giving
rise to a mechanism of saturation.

\begin{figure*}[t]
\centerline{\includegraphics[scale=1.0]{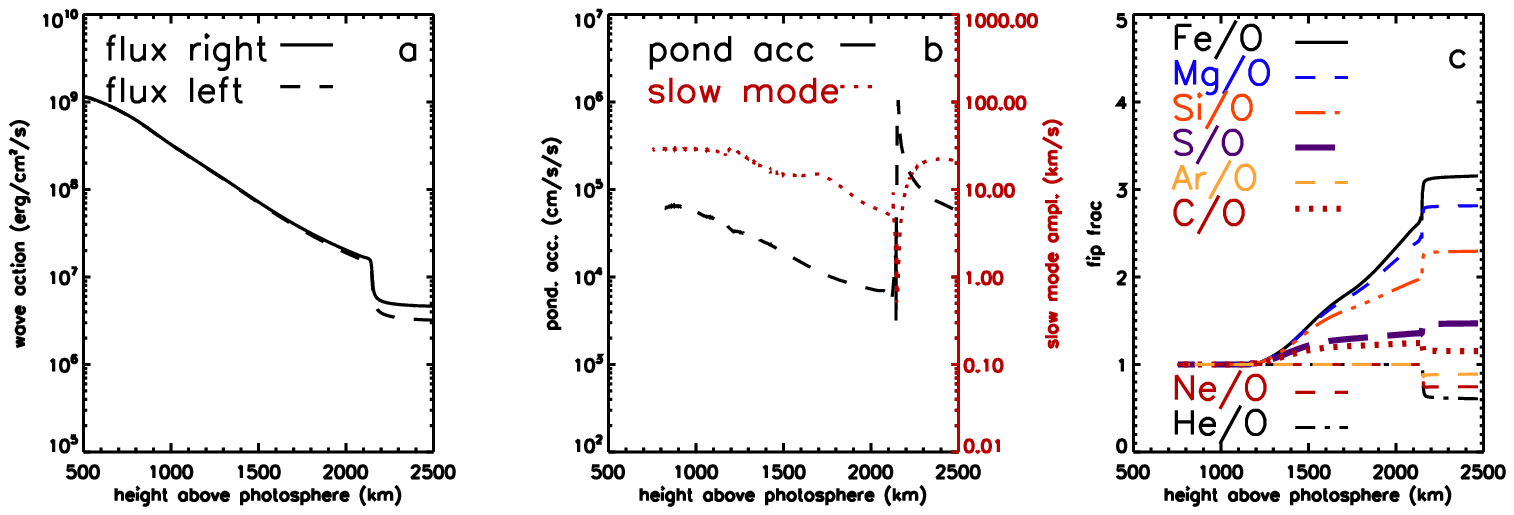}}
\centerline{\includegraphics[scale=1.0]{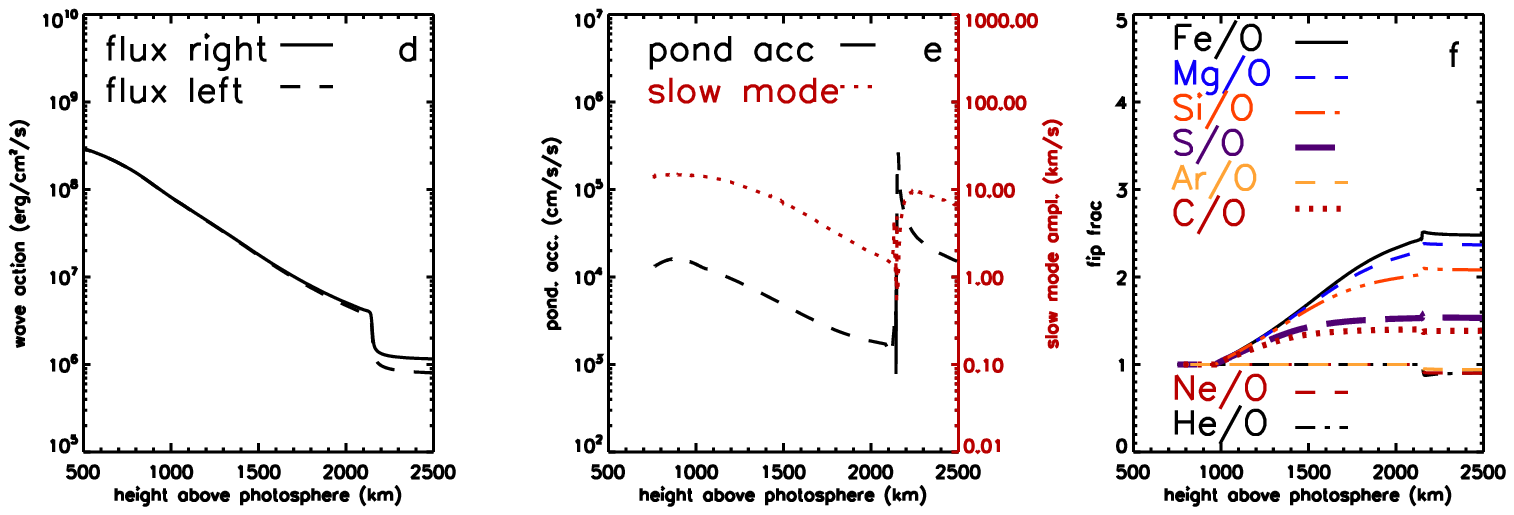}}
\centerline{\includegraphics[scale=1.0]{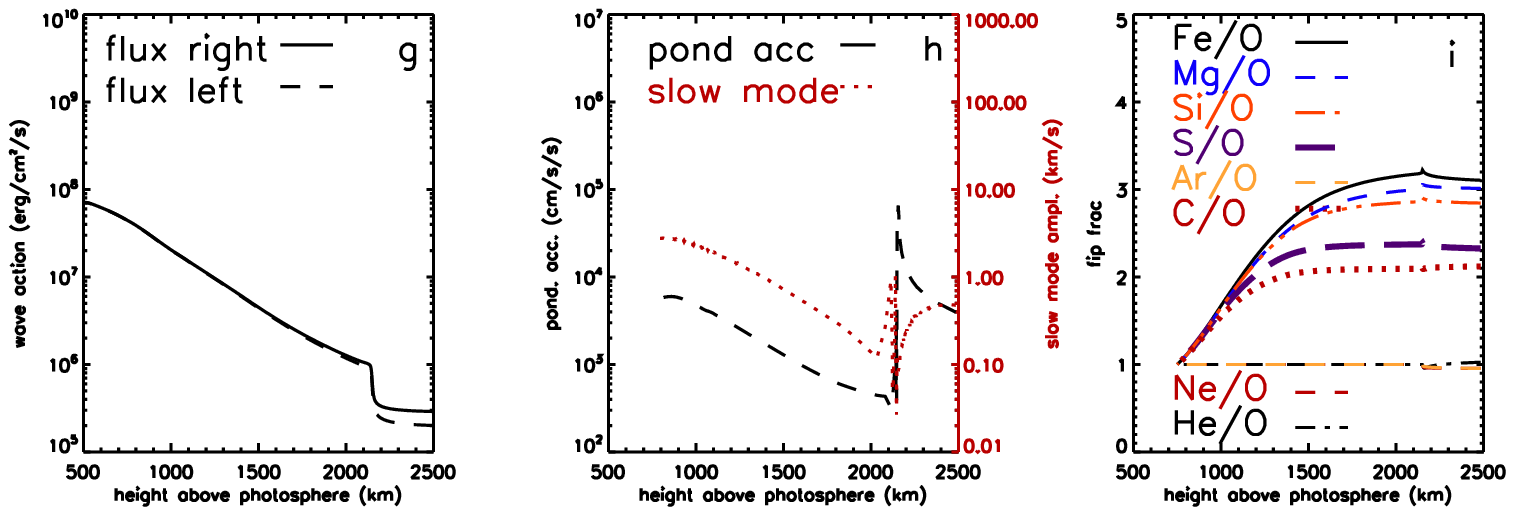}}
%\plotone{f5.ps}
\caption{\label{fig5}The chromospheric model for slow wind from an open field region
showing the varieties of FIP fractionation expected. Top row (a), (b), (c) show wave energy
fluxes, ponderomotive acceleration and slow mode wave amplitude, and fractionations
for high shear wave energy fluxes. Strong FIP fractionation, depletion of He, and small
enhancements of S and C result. Middle row (d), (e), (f) show the same plots for lower amplitude
shear Alfv\'en waves. Lower FIP fractionation, reduced He depletion, and stronger S
and C enhancements are seen. Bottom row (g), (h), (i) show the same plots for torsional Alfv\'en
waves. He depletion vanishes, even stronger fractionation of S and C are exhibited.}
\end{figure*}

\section{Fractionation Model Results}
\subsection{Coronal Hole}
Figure 3 shows the chromospheric portion of the solution of equations 4 for a
coronal hole. The magnetic field at the top of the chromosphere is 10G,
leading to a plasma $\beta =1$ layer at an altitude of 1,000 km above the
photosphere. The top panels show the chromospheric density and temperature
structure and the ionization balances for low and high FIP elements. The
bottom three panels show the energy fluxes (left and right going) for the
five shear Alfv\'en waves comprising the model spectrum, the ponderomotive
acceleration and amplitude of sound waves generated by the Alfv\'en wave
driver, and the resulting fractionations. At altitudes below about 1,400 km,
the amplitude of sound waves is higher than the local Alfv\'en speed and no
fractionation occurs. Above this height, fractionation of Fe, Mg, and Si sets
in with magnitude 1.5 - 2.0, S and C are much less enhanced, by 1.1 - 1.2,
and Ar, Ne and He are depleted. This depletion of He is characteristic of
fractionation concentrated at the top of the chromosphere.

\subsection{Closed Coronal Loop}
Figure 4 shows similar panels to Fig. 3, but now for a closed coronal loop
with a magnetic field at the top of the chromosphere of 30G. The $\beta =1$
layer is now at 750 km altitude. However fractionation still is only
significant at heights similar to that in the coronal hole. Here the reason
is that the model contains only one coronal shear Alfv\'en wave that is
trapped in the coronal loop, and insufficient wave energy leaks low enough in
the chromosphere to cause fractionation lower down. The pattern of
fractionation is similar, but larger than, that in the coronal hole. Fe, Mg,
and Si are enhanced by factors of 3 - 4.5, S and C by 1.1 - 1.6, and He and
Ar are again depleted. Similarly to the coronal hole, the ponderomotive
acceleration shows a spike of $\sim 10^6$ cm s$^{-2}$, about twice as large
as in the coronal hole, at an altitude of 2,150 km where the chromospheric
density gradient is strongest, and this is where the strongest fractionation
occurs.

\subsection{Slow Speed Solar Wind}
Both examples above show S and C behaving more like high FIP elements, in
that they do not fractionate appreciably. Figure 5 shows examples of
fractionation in open field with magnetic field similar to the closed loop
(30 G at the top of the chromosphere) that show a transition from S and C
behaving as high FIP elements as above, to becoming fractionated like the low
FIP elements. The top row, (a), (b), and (c) show the (shear) Alfv\'en wave
energy fluxes, and ponderomotive acceleration and associated slow mode
amplitude, and the FIP fractionations respectively. This example has
relatively high Alfv\'en wave amplitudes, leading to a spike in the
ponderomotive acceleration of about $10^6$ cm s$^{-2}$ as in the closed loop
case, and similar fractionation to that case. The slow mode wave amplitude
developing in the lower chromosphere is larger than the local Alfv\'en speed,
suppressing any fractionation there.

The middle row (d), (e), and (f), show similar plots, but for an open field
with lower energy fluxes for the shear Alfv\'en waves. The significance of
the spike in the ponderomotive acceleration is reduced, but slow modes lower
down are of lower amplitude, allowing fractionation to occur there. The shift
of FIP fractionation to lower altitudes where H is largely neutral allows S
and C to behave more like low FIP elements. In equation 3 we no longer have
$\nu _{ki}
>> \nu _{kn}$, and so a small departure of $\xi _k$ from unity no longer
suppresses the fractionation in the same way as it does in a background gas
of protons. Hence elements like S, P, and C can become fractionated low in
the chromosphere, whereas higher up they cannot. This behavior is even more
pronounced in the bottom row, panels (g), (h), and (i), which show a similar
model except that shear Alfv\'en waves have been replaced by torsional waves,
which generate even lower slow mode wave amplitudes
\citep{vasheghani11,laming16}. Even more fractionation occurs close to the
$\beta = 1$ layer, with correspondingly more S and C. Such waves, when
combined with mass dependent fractionation discussed in section 3 give the
green dashed curve in the middle panel of Fig. 1. Note the enhanced S, P, and
C compared to the magenta curve representing shear Alfv\'en waves. Such
effects are much less prominent in the closed loop model. Here shear and
torsional Alfv\'en waves produce essentially the same fractionation pattern
\citep{laming16}, because the Alfv\'en waves remain trapped in the loop and
do not penetrate to the lower chromospheric regions where H is neutral, and
where the amplitude of sound waves coming up from the photosphere is much
lower.

When FIP fractionation is concentrated low in the chromosphere, He remains
undepleted, and S, P, and C are fractionated. The reverse is true when FIP
fractionation is concentrated at the top of the chromosphere; He and Ne are
depleted, and S, P, and C are essentially unchanged. In this way the pattern
of FIP fractionation can be seen to be dependent on wave properties in the
solar atmosphere, and to offer an novel and unexpected window into this
physics. In particular, perturbations akin to torsional Alfv\'en waves are
expected as part of the slow solar wind release process through interchange
reconnection. Open field reconnecting with twisted closed field takes on the
twist \citep[e.g.][]{lynch14,higginson17} which propagates away. This process
should easily excite both upward and downward propagating torsional waves as
the it proceeds. Even so, this observed behavior may suppress the excitation
of sound waves compared to shear waves even more than that modeled in
\citet{vasheghani11} and \citet{laming16}, reinforcing our conclusion.

\section{Mass and Charge Dependent Fractionation and Acceleration}
\subsection{Introduction}
While FIP fractionation by the ponderomotive force is the dominant mechanism
of abundance modification, a number of other possibilities exist in the solar
wind. Analysis of solar wind samples returned by the Genesis mission has
revealed isotopic fractionation between fast and slow solar wind
\citep{heber12}, where lighter isotopes are more abundant relative to heavy
ones of the same element in the slow wind compared to the fast. This is the
reverse of what equation 3 would predict for the ponderomotive force, so
clearly other mechanism(s) must be at work.

\subsection{Inefficient Coulomb Drag}
Inefficient Coulomb Drag (ICD) is usually implemented following
\citet{geiss70}. Assuming that H flows fastest in the solar wind, the flow
velocity of other elements $v_k$ is calculated relative to that of H, $V_H$,
as
\begin{eqnarray}
\nonumber v_k &=& V_H-{3\sqrt{\pi}\over 4}{m_p\over 4\pi e^4\ln\Lambda}{k_{\rm B}T\over n}
\sqrt{{A+1\over A}{2k_{\rm B}T\over m_p}}\\
&&\times\left(V_H{dV_H\over dr}+
{GM_{\sun}\over r^2}\right)\left(2A-Z-1\over 2Z^2\right)\left(A+1\over A\right).
\end{eqnarray}
An important point is that abundance modifications are only sustainable in
the solar wind while there is a collisional connection back to the solar
disk. Once the flow becomes collisionless according to equation 1, no further
fractionation is possible. All elements passing through this region must
eventually flow out in the solar wind. And different elements become
collisionless at different altitudes, giving
\begin{equation}
v_k = V_H-{r\over v_w\left(r\right)}\left(V_H{dV_H\over dr}+
{GM_{\sun}\over r^2}\right)\left(2A-Z-1\over 2A\right),
\end{equation}
with the implication that in the solar wind ICD only fractionates particles
according to the variation of $r/v_w\left(r\right)$ where they freeze in,
since the factor $\left(2A-Z-1\right)/2A$ varies much less from ion to ion
than does $\left(2A-Z-1\right)/2Z^2$ in equation 6. At the time of writing,
this is a difficult effect to quantify, but we speculate that it results in
possibly a much smaller fractionation than quoted previously
\citep[e.g.][]{bodmer98,bochsler07}. The parameter controlling this most
closely will be the plasma density, which has the strongest variation with
$r$.

At 1 A.U. though, minor ions (including He) are generally observed to flow
faster than H \citep[e.g.][]{kohl06,berger11}, limiting the applicability of
equation 6. This preferential acceleration presumably sets in once the wind
has become collisionless, where equation 6 is no longer valid in any case.

\subsection{Gravitational Settling}
Gravitational settling in a closed coronal
loop prior to eruption can be modeled with the same equation. The whole loop
can be assumed to be collisionally coupled to the solar disk, so the
complication from the transition to collisionless plasma does not arise. From
the continuity equation
\begin{equation}
{\partial n_k\over\partial t}=-\nabla\cdot\left(n_kv_k\right)\simeq -2n_kv_k/L
\end{equation}
where $L$ is the loop length, and $v_k$ is the element settling velocity
(absolute magnitude) calculated from equation 6 with $V_H=0$. This has
solutions of the form
\begin{equation}
n_k\propto \exp\left(-2v_kt/L\right).
\end{equation}
Assuming $n\sim 10^9$ cm$^{-3}$, $T\sim 10^6$ K and $L=75,000$ km, the
gravitational settling ($1/\exp $) times evaluate to 1.5, 3.6, and 5.0 days
for He, O, and Ne respectively. Thus such abundance modifications are only
likely to occur in the most quiescent of solar coronal structures \citep[see
e.g.][]{raymond97}. \citet{landi15} observe variations in the Ne/O abundance
ratio consistent with this, in quiescent coronal streamers and in the slowest
speed solar wind at solar minimum of 2005 - 2008, with Ne/O increasing to
0.25 during this period, higher than its more usual value of 0.17.
\citet{kasper12} and \citet{rakowski12} observe the He/H and He/O abundance
ratios moving in the {\em opposite} direction. While He depletion can be
caused by the ponderomotive force as part of the FIP fractionation, Ne should
behave similarly. And the He depletions observed, He/H as low as 1\%
\citep{kasper12,kepko16}, appear to be too extreme to be reproduced by the
ponderomotive force, so gravitational settling where He settles relative to O
and O settles relative to Ne appears to be the most plausible explanation.
Finally, heavy ion dropouts are also observed on occasion in the solar wind
\citep{weberg12,weberg15}, clearly indicating gravitational settling prior to
plasma release into the solar wind as in \citet{feldman98} where Fe (settling
time according the above of 2.4 days) is seen depleted relative to Si (4.3
days).

\subsection{First Adiabatic Invariant Conservation}
In \citet{laming17}, it was argued that the dominant
mass-dependent fractionation effect should be that of the conservation of the
first adiabatic invariant, in conditions where the ion gyrofrequency $\Omega
=eB/m_kc >> \nu _{ip}$. When an ion undergoes many gyro-orbits around the
magnetic field line in the time between Coulomb collisions with other ions
(mainly protons), the magnetic flux enclosed by its orbit is conserved. Hence
$Br_g^2\propto v_{\perp}^2/B$ is constant ($r_g$ is the particle gyroradius),
giving rise to an acceleration
\begin{equation}
{dv_z\over dt}=-{1\over 2}{dB\over dz}{v_{\perp}^2\over B}
\end{equation}
when $v^2=v_z^2+v_{\perp}^2$ is constant. While the plasma is still
collisionally connected to the solar envelope (i.e. before it becomes
collisionless and undergoes acceleration into the solar wind) a mass
dependent fractionation results
\begin{equation}
f_a=\exp\left\{ -\int {{dB/dz}\left(v_{\perp}^2/B\right)\over
2k_{\rm B}T/m_k+v_{||,osc}^2+2u_k^2}dz\right\}.
\end{equation}
This arises because the thermal speeds $v_{\perp}^2$ and $2k_{\rm B}T/m_k$
are proportional to $1/m_k$, while $v_{||,osc}^2$ and $u_k^2$ representing
fluid motions are not, and are usually much larger, and can match the
isotopic differences between high speed and low speed solar wind.

\begin{table}
\begin{center}
\caption{Model Corona and Wind Fractionations}
\begin{tabular}{|c|cc|cc|cc|}
%{\textwidth}{|c @{\extracolsep{\fill}} |ccc|}
\hline%\noalign{\vspace{.5em}}
%Element & \multicolumn{2}{c}{Fast Wind}& \multicolumn{2}{c}{Closed Loop}
%& \multicolumn{2}{c}{Slow Wind}\\
Element & Closed& Loop& Slow& Wind& Fast & Wind\\
\hline
H & 0.81& 1.01& 1.39& 1.74& 1.01& 1.27\\
He& 0.57& 0.68& 1.03& 1.22& 0.74& 0.89\\
C & 1.16& 1.22& 2.12& 2.24& 1.07& 1.13\\
N & 0.84& 0.86& 0.98& 1.016& 0.87& 0.90\\
Ne& 0.75& 0.71& 0.96& 0.91& 0.79& 0.75\\
Na& 3.48& 3.20& 3.09& 2.84& 2.37& 2.117\\
Mg& 2.96& 2.68& 3.01& 2.73& 2.13& 1.93\\
Al& 2.79& 2.45& 2.95& 2.59& 1.98& 1.73\\
Si& 2.40& 2.08& 2.84& 2.47& 1.75& 1.52\\
P & 1.70& 1.43& 2.52& 2.12& 1.36& 1.14\\
S & 1.52& 1.26& 2.33& 1.94& 1.23& 1.02\\
Cl& 1.04& 0.84& 1.00& 0.80& 0.95& 0.77\\
Ar& 0.96& 0.74& 0.95& 0.74& 0.92& 0.71\\
K & 3.67& 2.86& 3.13& 2.44& 2.42& 1.88\\
Ca& 3.64& 2.81& 3.12& 2.41& 2.40& 1.85\\
Ti& 3.62& 2.61& 3.12& 2.25& 2.38& 1.71\\
Cr& 3.55& 2.48& 3.11& 2.17& 2.33& 1.63\\
Fe& 3.52& 2.39& 3.11& 2.11& 2.32& 1.58\\
Ni& 3.09& 2.05& 3.02& 2.01& 2.09& 1.39\\
Zn& 3.37& 2.15& 2.85& 1.824& 2.13& 1.36\\
\hline  %\noalign{\vspace{.5em}}

\hline
\end{tabular}

\end{center}
{\tablecomments{All fractionations given relative to O. First column for each
model gives ponderomotive fractionation, second column gives ponderomotive
and adiabatic invariant conservation combined, as shown in Fig. 1. Slow wind
model assumes torsional Alfv\'en waves.} \label{tab1}}
\end{table}

%\section{Resonant Ion-Cyclotron Heating and Solar Wind Acceleration}

\subsection{Resonant Heating by Ion Cyclotron Waves}
Since the advent of the SOlar and Heliospheric Observatory
\citep[SOHO][]{domingo95} and ensuing imaging missions, the solar atmosphere
has come to be increasingly appreciated as a dynamic and complex environment.
Waves play a much larger role in shaping the plasma properties than hitherto
assumed and they can have comparable energy densities to the thermal gas in
the corona. For example, a major discovery made by the Ultraviolet
Coronagraph-Spectrometer on SOHO \citep[UVCS][]{kohl95,kohl97,kohl06} was
that of significant heating in the O$^{5+}$ ion inferred from spectral line
broadening, beginning at altitudes where the plasma becomes collisionless
according to equation 1, and to a lesser extent also in Mg$^{9+}$. It is
likely that all heavy ions are heated in this manner and location--- O VI and
Mg X were the only ions accessible to UVCS observation with sufficient
counting statistics. This heating is believed to derive from resonance with
ion-cyclotron waves. Major questions surround the origin of the ion cyclotron
waves, with in situ generation, presumably via a turbulent cascade from lower
frequency Alfv\'en waves, being favored \citep{cranmer01,hollweg02}, and the
degree of isotropy in the heating, with strongly anisotropic energization
with perpendicular temperature, $T_{\perp}
>> T_{||}$, the parallel temperature, favored. This $T_{\perp}$ is converted
to parallel velocity in the expanding magnetic field lines by conservation of
the first adiabatic invariant, leading to solar wind acceleration.

Further insight into these processes can only come from observing ion
cyclotron resonant heating in a wider variety of ions, establishing the
spectrum of ion cyclotron waves and the rates of acceleration of various ions
into the solar wind. For example, ion energization can arise as ion cyclotron
waves progressively cascade to higher frequencies, or are brought into
resonance by frequency sweeping might be expected to lose all their energy to
the lowest gyrofrequency ion in the plasma \citep[e.g.][]{vocks02}, until the
velocity distribution function of that ion becomes sufficiently distorted to
reach marginal stability allowing wave to pass through that resonance to the
next lowest gyrofrequency ion. Such a case would have a quite different
distribution of ion non-thermal line broadenings to a case where ion
cyclotron waves were excited directly by e.g. reconnection \citep{liu11}.

\section{Observing Strategies to Test the Roles of Element Fractionation and MHD Waves}
\subsection{General Observing Concept}
Off-limb observations give the best view of the solar corona uncontaminated
by emission from plasma at lower altitudes. The choice of waveband is a
tradeoff between count rates and the selection of diagnostic lines available,
with the best compromise generally being found in the Far Ultraviolet (FUV)
and close by part of the Extreme Ultraviolet (EUV) wavebands. Higher
throughput may be achieved at longer wavelengths, especially from the ground
\citep[e.g. the Daniel K. Inouye Solar Telescope;][]{tritschler14}, but with
reduced availability of useful diagnostic lines for our specific purposes.
Pushing further into the EUV would give more useful lines, but with
diminished count rates due to mirror and grating reflectivities.
Additionally, the FUV/EUV combination includes the H Lyman series and also
lines with radiative and collisionally excited components, adding to the
diagnostic utility.

Such off-limb UV spectroscopy would directly observe the element abundance
fractionations (e.g. those illustrated in Table 1) in various coronal
structures, allowing these to be traced back to the solar disk and related to
the properties of MHD waves propagating in the solar atmosphere, with
particular references to how these waves interact and drive the solar wind
through ion cyclotron resonance. This approach drives the spatial and
temporal resolution of observations, discussed further below. Figure 6 shows
a schematic diagram of the observation concept. Slits observing off limb at
projected heliocentric distances between 1.3 and 3.0 $R_{\sun}$ return
Extreme UltraViolet (EUV) and Far UltraViolet (FUV) spectra. The slit heights
are chosen to represent the solar corona fluid-kinetic transition region
discussed above, where the acceleration of the solar wind commences, and a
region where solar wind acceleration and the associated line broadening
should be readily visible. Ideally, several slit configurations would be
available e.g. a single slit for detailed spectroscopy of the widest possible
selection of lines, and two slits for observing only the strongest lines for
wave and shock studies, allowing the discrimination between upward and
downward propagating, and standing waves.

In an alternative approach, the Spectral Investigation of the Coronal
Environment (SPICE) instrument on Solar Orbiter \citep{fludra13} views the
solar disk directly, in order to study the solar source of the wind
simultaneously detected {\it in situ} on the same spacecraft. It will view a
subset of the lines in our envisaged EUV bandpass, and use one slit (of
varying sizes) at a time. These observations will be more focussed on
identifying the precise sources of the solar wind through their abundance
patterns, and less on the wave physics and acceleration processes in the
extended corona. They will, however, have strong S lines within their
bandpass, allowing the study of some of the subtle fractionation issues
discussed above.

In the following subsections, we consider the spectral bandpasses in the FUV
and EUV which optimize the coverage of spectral lines from different low and
high FIP elements for FIP fractionation studies (5.2), and the special
considerations required for lines which are also excited radiatively by
absorption of light from the solar disk (5.3). Following those we discuss the
observing approach for abundance studies, specifically He and S (5.4), direct
wave observations with two slits (5.5), and the application of our strawman
instrumentation to other topics in solar wind science (5.6).

\begin{figure}[t]
\centerline{\includegraphics[scale=0.6]{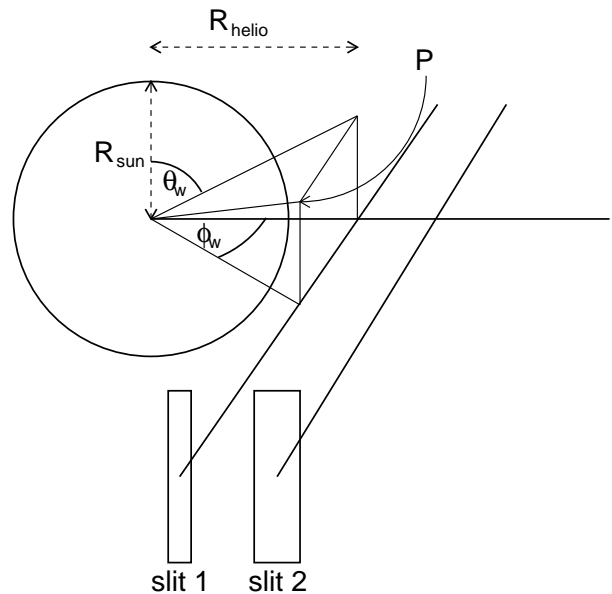}}

%\plotone{f5.ps}
\caption{Schematic of off-limb observation geometry. In this example, two
slits observe at heliocentric heights 1.5 and 2.2 $R_{\sun}$. Radiation on an
ion at point P at projected heliocentric radius $R_{helio}$ is calculated by
integrating across the line profile, then over the solar disk, and finally
along the line of sight.\label{fig6}}
\end{figure}

\subsection{Spectral Bandpass and Resolution}
In Tables 2 and 3 we give the spectral bandpasses (short and long in the UV
range, with count rates appropriate to the quiet solar corona) most
appropriate for testing the theoretical predictions above. They are similar
to the UVCS bandpasses, but with extended wavelength ranges to observe a
wider sample of coronal ions. The long wavelength region has been extended to
include the He II 1640 \AA\ multiplet. Lines from S X and S XI appear in both
long and short wavelength range. These become very important since the only
lines from carbon are C IV which are difficult to compare with other similar
temperature lines, and phosphorus has a low abundance making its lines
intrinsically weak. \citet{feldman97} identify the P IX 853.54 \AA\ and
861.10 \AA\ ($2s^2 2p^{3~4}S_{3/2} - 2s^22p^{3~2}P_{3/2,1/2}$), P XI 1307.57
\AA\ and 1317.66 \AA\ ($2s^2 2p^{3~4}S_{3/2} - 2s^22p^{3~2}D_{5/2,3/2}$) and
the P XII 1096.71 \AA\ $2s2p^{~1}P_1 - 2p^{2~1}D_2$ transitions, and
\citet{laming96} gives calculations of the density dependence of P IX
1317.66/1307.57 intensity ratio. Prominent lines from low FIP ions Mg VII,
VIII, IX, Si VII, VIII, IX, Fe X and XI are available in the short wavelength
EUV bandpass, while the long wavelength FUV region adds Mg IX, Fe XII, XIII.
High FIP ions are mainly available in the EUV; O VI, Ne VII, VIII, Ar VIII,
XII, with N V, O VII and Ar XI also present in the FUV.

The primary science discussed in this paper, that of measuring relative
element abundances as a means of understanding solar wind origins, does not
strongly constrain the required resolution, since spectral line intensities
are the main observables. Wave studies are more demanding in this respect.
The H I Lyman $\alpha$ line is typically 1 \AA\ wide
\citep[e.g.][]{laming13}, which suggests a minimum resolution of $\lambda
/\delta\lambda\sim 10^3$ (300 km s$^{-1}$). The flux given in Table 3 with an
effective area of $\sim 1$ cm$^2$ gives a count rate of 30 s$^{-1}$ in a $10
\times 100$ arcsec$^2$ region of the corona, allowing the accumulation of
$\sim 1000$ counts in 30 s. This in turn allows a determination of the line
centroid to $\sim 1/\sqrt{1000}\sim 0.03$ \AA\, or about 10 km s$^{-1}$.
Measurement of line profiles as a result of ion-cyclotron resonance heating
will require still higher resolution, of order 3000 to resolve a 100 km
s$^{-1}$ line broadening.
\begin{figure}[t]
\centerline{\includegraphics[scale=0.5]{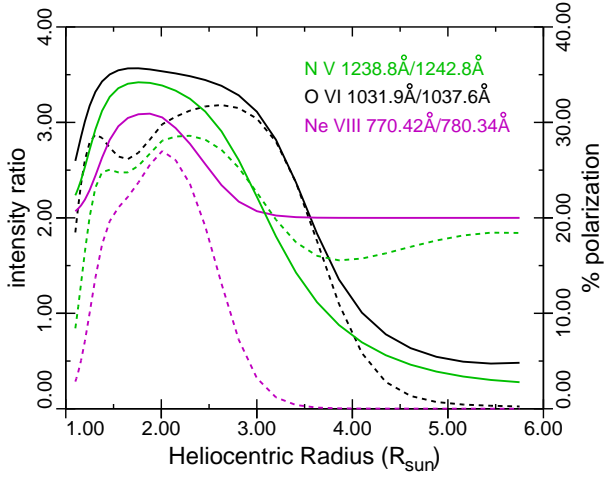}}

%\plotone{f5.ps}
\caption{N V, O VI and Ne VIII line intensity ratios as a function of
heliocentric radius, showing the intensity ratio (solid) and the polarization
resulting from radiative excitation (dash), to be read on the right-hand
axis. \label{fig8}}
\end{figure}

\begin{figure}[t]
\centerline{\includegraphics[scale=0.5]{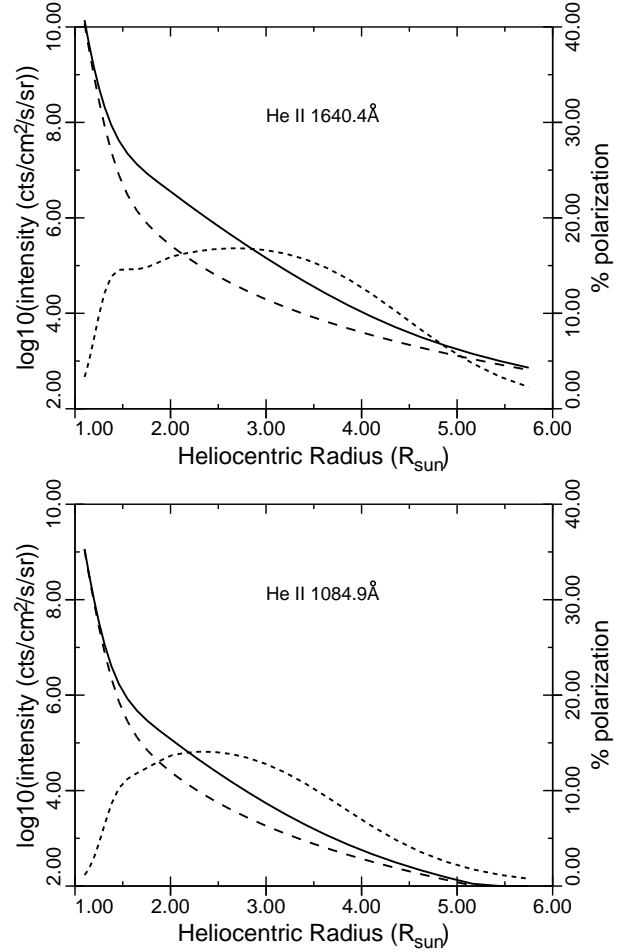}}

%\plotone{f5.ps}
\caption{He II line intensities as a function of heliocentric radius, showing
total (solid) and collisionally excited (long dash) components. The
radiatively excited component (not shown separately) is polarized, giving
overall line polarization shown by the short dash line to be read on the
right-hand axis.\label{fig7}}
\end{figure}

\begin{table*}
\begin{center}
\caption{Spectral Lines for Quiet Corona and Wind Fractionations, Short Wavelength}
\begin{tabular}{|l|l|l|l|l|}
%{\textwidth}{|c @{\extracolsep{\fill}} |ccc|}
\hline%\noalign{\vspace{.5em}}
%Element & \multicolumn{2}{c}{Fast Wind}& \multicolumn{2}{c}{Closed Loop}
%& \multicolumn{2}{c}{Slow Wind}\\
Wavelength & UVCS/SUMER & CHIANTI& Ion& Transition\\
\hline
  700.24&    5.5e7&        &    Ar VIII & $3s ^{~2}S_{1/2} - 3p^{~2}P_{3/2}$\\
  703.63&    1.6e7&        &     Al IX  & $2s^2 2p ^{~2}P_{3/2} - 2s 2p^{2~4}P_{3/}$\\
  706.05&    5.7e8&        &      Mg IX & $2s^{2~1}S_0 - 2s2p^{~3}P_1$\\
  713.81&    2.5e7&        &     Ar VIII&  $3s^{~2}S_{1/2} - 3p^{~2}P_{1/2}$\\
  749.55&    8.8e7&        &    Mg IX   &  $2s2p^{~1}P_1 - 2p^{2~1}D_2$\\
  770.42&    2.2e9&     6.6e9&    Ne VIII&   $2s^{~2}S_{1/2} - 2p^{~2}P_{3/2}$\\
  772.29&    1.0e8 &         &  Mg VIII &  $2s^2 2p ^{~2}P_{3/2} - 2s2p^{2~4}P_{5/2}$\\
  772.53&    6.0e6 &         &  Al VIII  &  $2s^2 2p^{2~3}P_2 - 2s 2p^{3~5}S_2$\\
  776.25&    1.70e8&        &   S X      &  $2p^{3~4}S_{3/2} - 2p^{3~2}P_{3/2}$\\
  780.34&    1.1e9 &    3.3e9&   Ne VIII  &  $2s^{~2}S_{1/2} - 2p^{~2}P_{1/2}$\\
  782.37&    8.0e7 &       &   Mg VIII  &  $2s^2 2p^{~2}P_{3/2} - 2s 2p^{2~4}P_{3/2}$\\
  782.96&    3.0e7 &        &   S XI     &  $2p^{2~3}P_1 - 2p^{2~1}S_0$\\
  789.44&    0.9e7 &        &   Mg VIII  &  $2s^2 2p^{~2}P_{3/2} - 2s 2p^{2~4}P_{1/2}$\\
  789.78&    0.4e7 &        &   Na VIII  &  $2s^{2~1}S_0 - 2s 2p ^{~3}P_1$\\
  854.66&    0.8e7&       &    Mg VII  &  $2s^2 2p^{2~3}P_1 - 2s 2p^{3~5}S_2 $\\
  868.11&    1.2e7&        &    Mg VII  &  $2p^{2~3}P_2 - 2p^{3~5}S_2$\\
  895.16&    1.2e8&     3.6e8&   Ne VII &   $2s^{2~1}S_0 - 2s2p ^{~3}P_1$\\
  944.37&    1.6e8&     4.8e8&   Si VIII & $2p^{3~4}S_{3/2} - 2p^{3~2}P_{3/2}$\\
  949.24&    5.4e7&     1.5e8&   Si VIII & $2p^{3~4}S_{3/2} - 2p^{3~2}P_{1/2}$\\
  950.16&    8.3e7&     2.5e8&   Si IX   & $2p^{2~3}P_1 - 2p^{2~1}S_0$\\
  972.54&    2.4e8&         &   H I     & $1s^{~2}S_{1/2}- 4p^{~2}P_{1/2,3/2}$\\
  1005.541&  5.5e5&   2.02e7&   Si VII & $2s^22p^3(^2P)3s^{~3}P_1 - 2s^22p^3(^2P)3p
  ^{~3}P_2$\\
  1009.908&     &   7.77e7&   Si VII & $2s^22p^3(^2P)3s^{~3}P_2 - 2s^22p^3(^2P)3p
  ^{~3}P_2$\\
  1018.60 &  2.5e7&        &   Ar XII & $2s^2 2p^{3~4}S_{3/2} - 2s^2
  2p^{3~2}D_{5/2}$\\
  1018.903&     &   4.64e7&   Fe XI  & $3s^2 3p^3 (^4S)3d{~5}D_3 - 3s^2 3p^3 (^2P)  3d
  ^{~3}F_4$\\
  1025.724&  1.2e9&   5.30e8&   H I    & $1s^{~2}S_{1/2} - 3p^{~2}P_{1/2,3/2}$\\
  1028.026&  5.0e7&   8.94e8&   Fe X   & $3s^2 3p^4 (^3P)  3d^{~4}D_{7/2} - 3s^2 3p^4 (^1D)  3d
  ^{~2}F_{7/2}$\\
  1028.957&  1.5e8&   2.47e8&  Fe XI   & $3s^2 3p^3 (^4S)  3d^{~5}D_4 - 3s^2 3p^3 (^2P)  3d
  ^{~3}F_4$\\
  1031.914&  4.0e9&   2.96e10&   O VI  & $1s^2 2s ^2S_{1/2} - 1s^2 2p
  ^2P_{3/2}$\\
  1037.615&  1.3e9&   1.48e10&  O VI   & $1s^2 2s ^2S_{1/2} - 1s^2 2p
  ^2P_{1/2}$\\
  1049.155&  7.0e6&   9.38e7&   Si VII & $2s^2 2p^{4~3}P_1 - 2s^2 2p^{4~1}S_0$\\
  1051.538&  1.0e6&   1.00e7&   Al VII & $2p^5 3s^{~3}P_2 - 2p^5 3p^{~3}S_1$\\
  1053.998&  3.0e6&   2.85e7&   Al VII & $2s^2 2p^{3~4}S_{3/2} - 2s^2
  2p^{3~2}P_{3/2}$\\
  1054.87 &  2.5e7&         &   Ar XII & $2s^2 2p^{3~4}S_{3/2} - 2s^2
  2p^{3~2}D_{3/2}$\\
  1056.917&  1.0e6&   1.14e7&   Al VII & $2s^2 2p^{3~4}S_{3/2} - 2s^2 2p^{3~2}P_{1/2}$\\
  1084.9  &      &   1.65e7 &   He II  &   2-5\\
  1132.774&  4.0e6&   3.94e7&   Si VII & $2s^2 2p^3 (^2D)  3s^{~3}D_3 - 2s^2 2p^3 (^2D)  3p
  ^{~3}F_4$\\
  1135.353&  1.6e7&   5.46e7&   Si VII & $2s^2 2p^3 (^4S)  3s^{~5}S_2 - 2s^2 2p^3 (^4S)  3p
  ^{~5}P_3$\\
  1137.240&  2.0e6&   1.57e7&   Si VII & $2s^2 2p^3 (^2D)  3s^{~3}D_2 - 2s^2 2p^3 (^2D)  3p
  ^{~3}F_3$\\
  1146.528&  2.0e6&   1.84e7 &  Si VII & $2s^2 2p^3 (^4S)  3s^{~5}S_2 - 2s^2 2p^3 (^4S)  3p
  ^{~5}P_1$\\
  1167.775&  2.4e6&   3.76e7&   Si VII & $2s^2 2p^3 (^4S)  3s^{~3}S_1 - 2s^2 2p^3 (^4S)  3p
  ^{~3}P_2$\\
  1182.455&      &   2.62e7&   Si VIII& $2s^2 2p^2 (^3P) 3s^{~4}P_{1/2} - 2s^2 2p^2 (^3P) 3p
  ^{~4}D_{3/2}$\\
  1183.995&  5.0e6&   6.13e7&   Si VIII& $2s^2 2p^2 (^3P) 3s^{~4}P_{3/2} - 2s^2 2p^2 (^3P) 3p
  ^{~4}D_{5/2}$\\
  1189.487&  1.4e7&   1.34e8&   Si VIII& $2s^2 2p^2 (^3P) 3s^{~4}P_{5/2} - 2s^2 2p^2 (^3P) 3p
  ^{~4}D_{7/2}$\\
  1189.867&  6.2e6&   6.72e7&   Mg VII & $2s^2 2p^{2~3}3P_1 - 2s^2 2p^{2~1}S_0$\\
\hline  %\noalign{\vspace{.5em}}

\hline
\end{tabular}

\end{center}
{\tablecomments{Intensities are in photons cm$^{-2}$ s$^{-1}$ sr$^{-1}$
computed from CHIANTI with a synthetic DEM matching that of the quiet Sun for
$\log T \ge 6.0$ with density = $10^7$ cm$^{-3}$. This is divided by 1000 to
match UVCS observations at $1.4 R_{\sun}$.} \label{tab2}}
\end{table*}

\begin{table*}
\begin{center}
\caption{Spectral Lines for Quiet Corona and Wind Fractionations, Long Wavelength}
\begin{tabular}{|l|l|l|l|l|}
%{\textwidth}{|c @{\extracolsep{\fill}} |ccc|}
\hline%\noalign{\vspace{.5em}}
%Element & \multicolumn{2}{c}{Fast Wind}& \multicolumn{2}{c}{Closed Loop}
%& \multicolumn{2}{c}{Slow Wind}\\
Wavelength & UVCS/SUMER & CHIANTI& Ion& Transition\\
\hline
  1196.217&  2.5e8&   3.61e8&   S X    & $2s^2 2p^{3~4}S_{3/2} - 2s^2
  2p^{3~2}D_{5/2}$\\
  1212.932&  5.0e8&   3.35e8&   S X    & $2s^2 2p^{3~4}S_{3/2} - 2s^2
  2p^{3~2}D_{3/2}$\\
  1215.670&      &   1.79e11&  H I    & $1s ^2S_{1/2} - 2p ^2P_{1/2,3/2}$\\
  1216.399&      &   2.04e7&   Si VIII& $2s^2 2p^2 (^3P) 3s^{~4}P_{3/2} - 2s^2 2p^2 (^3P) 3p
  ^{~4}D_{3/2}$\\
  1216.430&      &   2.33e8&   Fe XIII& $3s^2 3p^{2~3}P_1 - 3s^2 3p^{2~1}S_0$\\
  1238.823&  3.0e8&   8.70e8&   N V    & $1s^2 2s^{~2}S_{1/2} - 1s^2 2p
  ^{~2}P_{3/2}$\\
  1242.007&  1.0e9&   9.25e8&   Fe XII & $3s^2 3p^{3~4}S_{3/2} - 3s^2
  3p^{3~2}P_{3/2}$\\
  1242.806&  1.5e8&   4.36e8&   N V    & $1s^2 2s^{~2}S_{1/2} - 1s^2 2p
  ^{~2}P_{1/2}$\\
  1327.316&     &   1.24e7&   Mg VII & $2s^2 2p3s^{~3}P_1 - 2s^2 2p3p
  ^{~3}P_0$\\
  1334.223&      &   1.53e7&   Mg VII & $2s^2 2p3s^{~3}P_2 - 2s^2 2p3p
  ^{~3}P_2$\\
  1349.403&  5.0e8&   5.88e8&   Fe XII & $3s^2 3p^{3~4}S_{3/2} - 3s^2
  3p^{3~2}P_{1/2}$\\
  1350.439&      &   1.24e7&   Mg VII & $2s^2 2p3s^{~3}P_2 - 2s^2 2p3p
  ^{~3}P_1$\\
  1392.098&  1.7e8&   1.25e7&   Ar XI  & $2s^2 2p^{4~3}P_2 - 2s^2 2p^{4~1}D_2$\\
  1409.446&  3.0e7&   1.05e8&   Fe XI  & $3s^2 3p^3(^4S)  3d^{~5}D_3 - 3s^2 3p^3 (^2D)  3d
  ^{~1}G_4$\\
  1428.758&  1.5e8&   4.36e8&   Fe XI  & $3s^2 3p^3 (^4S)  3d^{~5}D_4 - 3s^2 3p^3 (^2D)  3d
  ^{~1}G_4$\\
  1440.510&  2.9e7&   2.53e9&   Si VIII& $2s^2 2p^{3~4}S_{3/2} - 2s^2
  2p^{3~2}D_{5/2}$\\
  1445.737&  3.6e8&   4.60e9&   Si VIII& $2s^2 2p^{3~4}S_{3/2} - 2s^2
  2p^{3~2}D_{3/2}$\\
  1463.489&  1.0e8&   7.85e8&   Fe X   & $3s^2 3p^4 (^3P)  3d^{~4}F_{9/2} - 3s^2 3p^4 (^1D)  3d
  ^{~2}F_{7/2}$\\
  1467.070&  1.3e8&   2.19e9&   Fe XI  & $3s^2 3p^{4~3}P_1 - 3s^2 3p^{4~1}S_0$\\
  1510.508&  &   2.87e7&   Ni XI  & $3s^2 3p^5 3d ^{~3}P_1 - 3s^2 3p^5 3d
  ^{~3}D_2$\\
  1537.282&      &   2.30e7&   Mg IX  & $2s 3s ^{~3}S_1 - 2s 3p ^{~3}P_2$\\
  1548.189&      &   4.48e8&   C IV   & $1s^2 2s ^{~2}S_{1/2} - 1s^2 2p
  ^{~2}P_{3/2}$\\
  1550.775&      &   2.24e8&   C IV   & $1s^2 2s ^{~2}S_{1/2} - 1s^2 2p
  ^{~2}P_{1/2}$\\
  1582.557&      &   3.30e8&   Fe XI  & $3s^2 3p^3 (^4S)  3d ^{~5}D_4 - 3s^2 3p^3 (^2D)  3d
  ^{~3}G_5$\\
  1603.209&      &   5.85e8&   Fe X   & $3s^2 3p^4 (^3P)  3d ^{~4}D_{7/2} - 3s^2 3p^4 (^1D)  3d
  ^{~2}G_{7/2}$\\
  1603.351&      &   2.81e8&   Fe X   & $3s^2 3p^4 (^3P)  3d ^{~4}D_{5/2} - 3s^2 3p^4 (^1D)  3d
  ^{~2}G_{7/2}$\\
  1604.779&      &   5.93e7&   Al VII & $2s^2 2p^{3~4}S_{3/2} - 2s^2
  2p^{3~2}D_{3/2}$\\
  1605.938&      &   6.06e7&   Ni XI  & $3s^2 3p^5 3d ^{~3}P_2 - 3s^2 3p^5 3d
  ^{~1}F_3$\\
  1611.710&      &   2.03e8&   Fe X   & $3s^2 3p^4 (^3P)  3d ^{~4}D_{7/2} - 3s^2 3p^4 (^1D)  3d
  ^{~2}G_{9/2}$\\
  1614.390&      &   7.03e7&   Fe XI  & $3s^2 3p^3 (^4S)  3d ^{~5}D_3 - 3s^2 3p^3 (^2D)  3d
  ^{~3}G_4$\\
  1614.495&      &   2.09e7&   S XI   & $2s^2 2p^{2~3}P_1 - 2s^2 2p^{2~1}D_2$\\
  1623.609&      &   7.51e7&   O VII   & $1s 2s ^{~3}S_1 - 1s 2p ^{~3}P_2$\\
  1639.777&      &   6.24e8&   Fe XI  & $3s^2 3p^3 (^4S)  3d ^{~5}D_4 - 3s^2 3p^3 (^2D)  3d
  ^{~3}G_4$\\
  1639.861&      &   1.49e7&   O VII  & $1s 2s ^{~3}S_1 - 1s 2p ^{~3}P_0$\\
  1640.40 &      &   1.40e8 &   He II  &   2 - 3\\
\hline  %\noalign{\vspace{.5em}}

\hline
\end{tabular}

\end{center}
{\tablecomments{Intensities are in photons cm$^{-2}$ s$^{-1}$ sr$^{-1}$
computed from CHIANTI with a synthetic DEM matching that of the quiet Sun for
$\log T \ge 6.0$ with density = $10^7$ cm$^{-3}$. This is divided by 1000 to
match UVCS observations at $1.4 R_{\sun}$.} \label{tab3}}
\end{table*}

\subsection{Radiative Excitation}
Also shown  in Fig. 6 is the geometry for computing the radiative excitation
component of lines which are illuminated by disk radiation. For calculations
where a detailed line profile is required, as in \citet{laming13}, the
radiatively excited component is calculated as a four dimensional nested
integral, integrating over the frequency overlap between the disk and coronal
line profiles, the azimuthal and poloidal angles, $\phi$ and $\theta$
respectively from point P projecting back to the solar disk, and finally the
distance along the line of sight. The range for $\phi$ is
\begin{eqnarray}
\nonumber &\phi_w-\sqrt{\left(\arccos\sqrt{1-1/R^2}\right)^2+\left(\theta
_w-\theta\right)^2}\le\phi\le \\
&\phi_w-\sqrt{\left(\arccos\sqrt{1-1/R^2}\right)^2-\left(\theta
_w-\theta\right)^2},
\end{eqnarray}
and for $\theta$;
\begin{equation}
\theta
_w-\arccos\sqrt{1-1/R^2}\le\theta\le\theta _w+\arccos\sqrt{1-1/R^2},
\end{equation}
where $R=R_{helio}\sqrt{\sec ^2\phi _w +\cot ^2\theta _w}$ is the
heliocentric distance of point $P$. For applications where a detailed line
profile is not required and predicted intensities are sufficient, and where
the coronal ion distribution can be assumed isotropic and the solar
illumination uniform, the integration over angles can be replaced by
multiplying by the solid angle $2\pi \left(1-\sqrt{1-R^2_{\sun}/R^2}\right)$.
\citet{auchere05} investigated relaxing both of these idealizations, and
\citet{raouafi04} considered deviations from radial flow in the solar wind
caused by the super-radial expansion of magnetic field lines, though further
discussion is beyond our scope here.

This last approximation is used in calculating the radiative and collisional
components of the Li-like doublets, N V, O VI and Ne VIII shown in Fig. 7,
and of the He II 1640.4 and 1084.9 \AA\ multiplets, shown in Fig. 8, both for
quiet Sun conditions. In Fig. 7, the solid curves give the intensity ratio
between the two components of the doublet. In conditions of pure collisional
excitation, this is precisely 2. Radiative excitation favors the stronger of
the two components, so the intensity rises above 2 as we move off-limb, with
a theoretical maximum in conditions of pure radiative excitation of 4. Above
about $3-4 R_{\sun}$, the coronal lines are Doppler shifted out of resonance
with the disk emission (known as ``Doppler dimming'') , and the intensity
ratio returns to 2, unless other lines exist in the disk spectrum (e.g. Fe
XII 1242.007 \AA\ in the case of N V and C II 1036.34/1037.02 \AA\ for O VI;
see Li et al. 1998) which move into resonance to continue the radiative
excitation. The short dashed lines show the (negative, i.e. orthogonal to the
radial direction) polarization in the stronger component of the doublet,
arising due to the radiative excitation \citep[e.g.][]{cranmer98}.

In Fig. 8 in each case the solid line gives the total He II line intensity.
The long dash lines give the contribution from collisional excitation, the
remainder being radiatively excited by emission from the solar disk in the
$1s-3p$ 256.37 \AA\ and the $1s-5p$ 237.36 \AA\  lines for 1640.4 and 1084.9
\AA\ respectively. These last processes are calculated using line intensities
observed by \citet{malinovsky73} and \citet{linsky76}, with line widths
inferred from \citet{brown08}. Close to the solar limb collisional excitation
dominates. As the electron density declines moving out into the corona,
collisional excitation proportional to density squared declines faster than
radiative excitation, and radiative excitation dominates. Even further out,
beyond 3-4 $R_{\sun}$ -- in this example, the acceleration of the solar wind
has Doppler-shifted the coronal line profile out of coincidence with the disk
emission, and collisional excitation is again important. Radiative excitation
leads to linear polarization in the line, given by the short dash curve to be
read on the right-hand axis.

The polarization has two effects on the measured line intensity, compared to
the unpolarized case. The first is that the polarized light is emitted
anisotropically. This is included in the calculation using the redistribution
functions given by \citet{cranmer98}. The second effect arises if the
instrumentation has polarization sensitivity, which needs to be corrected
for. In the usual case, to be discussed further below, using gratings near
normal incidence, (the grating is the dominant polarization sensitive
component), we estimate a polarization sensitivity of only a few \%, which
when observing a line polarized to 10-20 \% leads to errors in the intensity
measurement of order 1\%.  This is well below other uncertainties (mainly
counting statistics), and so is considered negligible from here on. A future
experiment to measure the polarization in the He II lines could be directly
interpreted in terms of the acceleration of the He component of the solar
wind.

The coronal He abundance (Laming \& Feldman 2001; 2003) is also a key
diagnostic of solar wind acceleration. Rakowski \& Laming (2012) showed that
He abundance variations also result from the ponderomotive force that
generates the FIP fractionation. Kasper et al. (2007; 2012) find extreme He
abundance variations in the slowest speed solar wind near solar minimum.
There is a complex interplay between the heating, acceleration and wave
absorption by helium (e.g., Kasper et al. 2008; Bourouaine et al. 2011, 2013;
Chandran et al. 2013; Verscharen et al. 2013) and direct observation of such
complex interplay will provide strong confirmation of wave driving of the
solar wind.
%UVCS observations suggested that the waves
%in question are generated in the extended corona near where they are damped
%(Cranmer 2000, 2001), but further observations are necessary.

\begin{table*}
\begin{center}
\caption{Wave Modes Determined by Correlations between Oscillations of $\delta$I, $\delta$W,
and $\delta\lambda$.}
\begin{tabular}{|p{0.5in}|p{1.25in}|p{1.25in}|p{1.25in}|p{1.25in}|}
%{\textwidth}{|c @{\extracolsep{\fill}} |ccc|}
\hline%\noalign{\vspace{.5em}}
 & \multicolumn{2}{c}{Line shift $\delta\lambda =0$}&
\multicolumn{2}{c}{Line shift $\delta\lambda \ne 0$}  \\
\hline
 & Line width $\delta W=0$& Line width $\delta W\ne 0$ &
 Line width $\delta W=0$& Line width $\delta W\ne 0$  \\
\hline
 Intensity $\delta I=0$ & Shear or torsional Alfv\'en wave with p.o.s. oscillation
 ({\bf k} along l.o.s) & Unresolved torsional Alfv\'en with {\bf k} in p.o.s.& Shear
 Alfv\'en wave with {\bf k} in p.o.s.& (Partially) resolved torsional Alfv\'en wave with {\bf k} in p.o.s.\\
 \hline
 Intensity $\delta I\ne 0$&No wave & Sound wave or fast mode with oscillation in p.o.s.& No wave
 & Sound wave or fast mode with oscillation along l.o.s.\\
\hline
\end{tabular}
\end{center}
{\tablecomments{p.o.s. = plane of sky; l.o.s. = line of sight} \label{tab4}}
\end{table*}

\subsection{Abundances and Waves}
The discussion above suggests that coronal sources of the slow speed solar
wind may be detectable by their abundance signature(s). Table 1 summarizes
the fractionations expected for fast wind, closed coronal loop, and slow wind
(i.e. open field, but with similar magnetic field to the closed loop), in
each case for ponderomotive fractionation alone, and for a combination of
ponderomotive and adiabatic invariant conservation designed to reproduce the
isotopic fractionation seen in Genesis sample return data. While the basic
FIP fractionation can be similar between material originating in closed loops
or in open field region, subtle details like the fractionation of S, P and C,
and also He and Ne can vary. {\em This is potentially an important
diagnostic.} The slow solar wind is believed to originate in interchange
reconnection between closed and open field, and the released wind should have
have a composition determined by the relative amounts of originally closed
and open field plasma that ultimately are released. The S abundance
measurements of \citet{giammanco07a,giammanco07b} suggest that this is indeed
the case, falling as they do between our closed loop and open field slow
solar wind models. We therefore expect He, C, Ne, P and S, the elements that
change the most between the closed loop and slow wind models in Table 1, to
be the most important element abundances to study.

To estimate the potential instrument performance, let us consider the
requirement for detecting a factor of two change in the S or He abundance.
The most intense lines of S are the S X 1196.217 and 1212.932 \AA\ lines in
the FUV bandpass, where the effective area is usually highest. Assuming a
10'' slit at 1.5$R_{\sun}$, the solar wind is moving at 20 km s$^{-1}$ and
takes 375 s to cross the slit field of view. Taking 100 counts (very
conservative) in each line as a minimum to detect a factor of two change, and
an effective area of 0.2 cm$^2$, this can be done in 375 s within a solid
angle of $3.7\times 10^{-9}$ rad, or approximately 12.5''$\times 12.5$''. The
He II 1640.4 \AA\ multiplet is less intense, and falls in a region of lower
throughput, leading to about an order of magnitude less signal, or meaningful
abundance measurements when integrated over 10'' $\times 100$'' solid angle.
At 2$R_{\sun}$ heliocentric distance, the line intensities are two orders of
magnitude lower. The effective area, however, could be higher and a wider
slit would lead to about one order of magnitude higher count rate.

%Although compressive MHD waves connected to slow mode or fast mode waves, or
%their analogs in inhomogeneous media, have been readily observed (e.g.,
%Aschwanden et al. 1999; Nakariakov et al. 1999), the less obvious
%noncompressive Alfv\'en waves are also beginning to be detected (Tomczyk et
%al. 2007). And as van Dooresselaere et al. (2008) emphasize, the new
%realization that Alfv\'en or fast mode waves (loosely collectively referred
%to as ``Alfv\'enic'' when close to parallel propagation) are ubiquitous in
%the solar upper atmosphere (McIntosh et al. 2011) signifies an important new
%development with profound consequences for our understanding of the corona
%and solar wind.

\subsection{Direct Wave Observations with Two Slits}
The different regimes of the solar wind are known to be distinguishable by
their turbulence and wave characteristics \citep[e.g.][]{bruno13,ko18}. The
fast wind shows mainly Alfv\'enic, but unbalanced turbulence, while the slow
wind is more balanced, but less Alfv\'enic. These characteristics match
naively with the thought that FIP fractionated slow wind originates in closed
coronal loops, where the balanced turbulence is a relic of that trapped in
the loop, while relatively unfractionated fast wind originates from open
field where, with Alfv\'en waves propagating up from the chromosphere, the
turbulence is naturally unbalanced.

However the solar wind is not so simple. Interchange reconnection between
open and closed field is necessary to allow the plasma originally contained
in loops to escape \citep[e.g.][]{antiochos11}. And as we have shown above
(subsection 2.6), the high S abundance in the slow speed solar wind appears
to require nonresonant waves, most plausibly from an open field region. The
double slit geometry outlined above would allow us to make direct
observations of waves, and assess their frequency, wavenumber, mode,
cross-helicity, etc.

Table 4 summarizes the observational properties of the various MHD wave
modes, and how they might be identified from variations in line centroid,
width and intensity. Simultaneous observations in two slits also allows
inferences on direction of motion and cross-helicity (i.e. degree to which
waves are ``balanced''). Consider two counterpropagating waves with
amplitudes $a$ ad $b$,
\begin{eqnarray}
\nonumber &a\exp i\left(\omega t -kz\right) + b\exp i\left(\omega t+kz\right)=\\
&2b\exp i\omega t\cos kz +\left(a-b\right)\exp i\left(\omega t -kz\right).
\end{eqnarray}
With the first slit at $z=0$, the signal is $\propto \left(a+b\right)\exp
i\omega t$ but at a second slit a projected distance $L$ away, the signal is
$2b\exp i\omega t\cos kL +\left(a-b\right)\exp i\left(\omega t-kL\right)$.
The balanced portion of the disturbance produces oscillations is in phase in
both slits. The unbalanced portion produces a second oscillation in the
second slit with phase difference $-ikL$. Except when $kL=2\pi$, balanced and
unbalanced waves can be diagnosed, for comparison with predictions coming
from the abundance pattern.

Low-frequency coronal waves themselves will be thus revealed by careful
observation of the central region of the Ly$\alpha$ line profile. Recently,
5-minute Alfv\'enic waves have been detected with the Coronal Multichannel
Polarimeter (CoMP; Tomczyk et al. 2007) at low heights where the plasma is
collisional, together with Alfv\'enic turbulence in coronal loops (De Moortel
et al. 2014; Liu et al. 2014), and in open field (Morton et al. 2015), albeit
with lower amplitudes than expected. McIntosh \& De Pontieu (2012) discuss
possible reasons for this, e.g., the ``dilution'' of the signal by foreground
and background emission. It is also possible that waves exist with higher
amplitudes at different frequencies, as yet undetected. Mancuso \& Raymond
(2015) detect propagating kink waves (an almost parallel propagating fast
mode wave) with SOHO/UVCS revealed by Doppler shift oscillations in H I
Ly$\alpha$.

\subsection{Shock Waves} Coronal Mass Ejections (CMEs) can also drive waves
through the solar corona with important consequences for SEP acceleration
when these steepen into shocks as the magnetic field decreases off-limb.
Two-slit observations can determine the height of shock formation and the
plasma properties of the pre-CME corona (Raymond et al. 2003). It is
important to correlate the He abundance of the pre-CME corona with the large
variations of He abundance in SEPs. Limits on the Alfv\'en and shock speeds
(key parameters in SEP acceleration models) can be set by detection of the
shock arrival at different heights as determined by the timing of the
increase in line widths of UV emission lines (Mancuso et al. 2003; Ciaravella
et al. 2006). The angle between the shock front and the magnetic field
requires the pre-shock field direction, which can be determined from streamer
morphology.

Shocks are seen in white light images because they compress the plasma
\citep{vourlidas03,vourlidas13,liu17}. They appear in UV spectra as drastic
increases in line widths due to shock heating \citep{mancuso02,vourlidas12}.
These observations provide the compression ratio in the shock \citep[a key
parameter determining SEP spectral shape, e.g.][]{kwon18} and information
about thermal equilibration among electrons, protons and ions (Bemporad et
al. 2014). At high effective area, a large number of ionization states will
be available for observation with this instrument concept, revealing the
progress of ionization behind the shock consistently and providing the
electron temperature (Ma et al. 2011). UV spectroscopy can test collisionless
theories of multi-ion shock heating as a function of mass-per-charge (Lee \&
Wu 2000; Zimbardo 2011). Polarimetry, if available, can also yield inferences
on shock microphysics \citep{shimoda18}, following the initial prediction
\citep{laming90} and discovery \citep{sparks15} of polarized emission in
H$\alpha$ from collisionless shock waves in SN 1006. In the solar case, H I
Lyman $\alpha$, usually polarized in a North-South direction due to resonant
scattering of disk radiation, will be Doppler shifted out of resonance with
the disk line and become polarized in a direction along the shock velocity
vector, usually close to East-West, by collisions with the anisotropic
post-shock electron and proton distributions.

Spectroscopy can also shed light into the heating and acceleration of CMEs.
UVCS observations of Fe$^{17+}$ ($T_e\simeq 6$ MK) within thin structures
trailing CMEs, provide evidence for reconnection in current sheets, a key
prediction of many CME initiation models (Ciaravella et al. 2002; Lin et al.
2015). Other UVCS measurements show that the thermal energy is comparable or
can even exceed the CME kinetic energy (Akmal et al. 2001; Murphy et al.
2011). Yet, these observations are few and far between to allow a detailed
investigation of the energy transfer in eruptive events. An instrument
concept with greatly increased sensitivity over the UVCS telescope will
observe many high temperature (multiple Fe ions from 18+ to 21+) and
low-temperature lines (C 3+, Si 3+) that can greatly expand our understanding
of the CME initiation and initial evolution.

\section{Conclusions}
Our emerging understanding of FIP fractionation in terms of the ponderomotive
force due to Alfv\'en waves, and improved observations revealing hitherto
unexpected variations in the abundances of He, S, P, and C, suggest that we
are on the cusp of significant breakthroughs in solar wind science. The S, P,
C abundance enhancements can be traced to the differing altitudes in the
chromosphere at which fractionation occurs, and this in turn can be traced to
the differing properties of the Alfv\'en waves causing the fractionation,
with respect to the magnetic structures in which they are propagating.
Relationships should exist between the solar wind abundances and the nature
of the turbulence entrained within it, a prime example being the
cross-helicity or degree of balance between sunward and anti-sunward
propagating waves. The cross-helicity is a crucial parameter in the
development of a turbulent cascade, by means of which fluctuations on large
scales can be transferred to smaller and smaller scales until they resonate
with solar ion winds, heating and ultimately accelerating them.

Multi-slit off-limb spectroscopy in the EUV and FUV thus holds great promise
for discoveries in solar wind science. Following on from the pioneering
observations of SOHO/UVCS, with modern fabrication techniques we expect an
increase of over a factor of 100 in instrument sensitivity, greatly extending
the range of detectable spectral lines and the height off-limb at which
observations can be made. Extending the UVCS bandpass to include the He II
1640 \AA\  multiplet will capture He abundance variations, as well as S and
C. Solar wind acceleration is one of the phenomena associated with the
transition from fluid to collisionless plasma, and it offers a probe of this
third transition layer in the solar atmosphere.

\acknowledgements This work was supported by grants from the NASA
Heliophysics Supporting Research (NNH16AC39I), Chandra Theory
(MIAA1702-0002-00), Heliophysics Grand Challenges (NNH17AE96I) and the
Laboratory Analysis of Returned Samples Programs (NNH17AE60I), and by basic
research funds of the Chief of Naval Research. JML also acknowledges the
hospitality of the International Space Science Institute in Bern where some
of this work was started. AV was supported by NASA HSR grant NNX16AG86G and
internal JHUAPL funds.

\end{document}